# Rapid and Precise Determination of Zero-Field Splittings by Terahertz Time-Domain Electron Paramagnetic Resonance Spectroscopy

Jian Lu[1], I. Ozge Ozel[2], Carina Belvin[2], Xian Li[1], Grigorii Skorupskii[1], Lei Sun[1], Benjamin K. Ofori-Okai[1], Mircea Dincă[1], Nuh Gedik[2], and Keith A. Nelson[1,*]

[1]*Department of Chemistry, Massachusetts Institute of Technology, Cambridge, Massachusetts 02139, USA*

[2]*Department of Physics, Massachusetts Institute of Technology, Cambridge, Massachusetts 02139, USA*

\* kanelson@mit.edu.

**Abstract:** Zero-field splitting (ZFS) parameters are fundamentally tied to the geometries of metal ion complexes. Despite their critical importance for understanding the magnetism and spectroscopy of metal complexes, they are not routinely available through general laboratory-based techniques, and are often inferred from magnetism data. Here we demonstrate a simple tabletop experimental approach that enables direct and reliable determination of ZFS parameters in the terahertz (THz) regime. We report time-domain measurements of electron paramagnetic resonance (EPR) signals associated with THz-frequency ZFSs in molecular complexes containing high-spin transition-metal ions. We measure the temporal profiles of the free-induction decays of spin resonances in the complexes at zero and nonzero external magnetic fields, and we derive the EPR spectra via numerical Fourier transformation of the time-domain signals. In most cases, absolute values of the ZFS parameters are extracted from the measured zero-field EPR frequencies, and the signs can be determined by zero-field measurements at two different temperatures. Field-dependent EPR measurements further allow refined determination of the ZFS parameters and access to the *g*-factor. The results show good agreement with those obtained by other methods. The simplicity of the method portends wide applicability in chemistry, biology and material science.

## Introduction

Transition-metal or rare-earth molecular complexes and biological molecules assume well-defined symmetries, and may undergo structural distortion to lower symmetries due to interactions of the central metal ions with each ligand.[1] The molecular orbitals consequently undergo crystal field splitting, and valence electrons rearrange to form an energetically stable ground state. Zero-field splitting (ZFS) refers to the magnetic sublevel fine structure of unpaired electrons in such molecular orbitals in the absence of an external magnetic field. ZFS originates from spin-spin interactions mediated by the ligand field and from spin-orbit coupling.[1] The behavior of ZFS can be adequately captured by two parameters, *D* and *E*, which are the axial and transverse components of the magnetic anisotropy, respectively.[2]

Accurate determination of the ZFS parameters for atomic centers with unpaired spins is critical in many research fields. For instance, in nanoscale thermometry based on nitrogen vacancy centers in diamond, the temperature dependence of *D*, the axial magnetic anisotropy, enables precise measurements of local temperature[3]. In metalloproteins, ZFS parameters are essential for interpreting EPR signals and understanding the electronic structure of metallo-cofactors[4]. In molecular magnets, *D* and *E* are related to both magnetic susceptibility and magnetization[5]; measuring these parameters independently anchors the structure-function relationship that enables the design of molecules with higher effective barriers for spin inversion. Indeed, in such molecules, a negative *D* parameter indicates Ising-type magnetic anisotropy, and the potential energy surface of the spin sublevels has a double-well shape, with a spin inversion barrier that is proportional to *D*.[6] Due to their long relaxation and coherence times at low temperatures, single-molecule and single-ion magnets have been proposed to serve as the smallest units in high-density memory, quantum information processing and spintronics,[7] as such systems have two stable states that can be switched reversibly by magnetic fields, analogous to nitrogen vacancy centers in diamond[8].

Conventional pulsed EPR spectroscopy using microwave technology has been widely applied to measure ZFS parameters between 1 and 100 GHz in the time domain. However, ZFS parameters of many metalloproteins and molecular complexes, in particular molecules with high magnetic anisotropy, lie in the THz frequency range between 0.1 and 10 THz (1 THz = 33.3 cm$^{-1}$). Access to these higher frequencies is not available currently with routine magnetic resonance techniques. In high-frequency, high-field EPR (HFEPR), an external magnetic field is continuously swept to shift the spin resonances to the frequencies of the narrowband THz radiation sources used.[9,10] Leading work in this area is conducted with magnetic fields that typically can be varied continuously from 0 to 25 T.[11] Compounds with particularly large values of the ZFS parameters would require even higher magnetic fields which scale up with increasing ZFS parameters[12]. Because these measurements are conducted with applied magnetic fields, they also do not allow for direct measurements of the ZFS parameters, which are instead inferred from fitting the field-swept resonance data at discrete frequencies. Alternatively, frequency-domain Fourier transform EPR using THz sources from coherent synchrotron radiation[13] or blackbody radiation[14,15] has been used to measure ZFS parameters in single-molecule magnets[13,16] and biological systems[14,17] at zero and nonzero external magnetic field. Although this allows direct and broadband measurements of ZFS parameters and enhanced precision when the field-swept broadband spectra are fitted by theoretical modeling[18], the synchrotron THz sources employed thus far place severe limitations on wide applicability in the community. Inelastic neutron scattering has been used to directly measure ZFSs, but a typical experiment requires large amounts of sample (often more than 1 g), and is only available at specialized facilities with neutron sources. Most commonly, temperature-dependent magnetometry measurements have been used to determine ZFS parameters. Although more widely available, this technique is also arguably the most imprecise, because the measurements inevitably convolve ZFS parameters with other magnetic responses (e.g., exchange coupling, which is often of similar frequency to ZFS) and average over the temperature variation of ZFS values. Indeed, ZFS parameters determined by magnetometry often deviate from those measured by other methods. Clearly, the development of a simple technique employing benchtop equipment to directly and reliably measure absolute values of ZFS parameters could have a transformative effect in physical inorganic and organic chemistry, leading to more facile development of new magnetic materials and greatly facilitating the understanding of structure and function in metalloenzymes.

In this work, we present direct characterization of THz-frequency ZFSs in transition-metal complexes by measuring their EPR spectra using THz time-domain spectroscopy both at zero and nonzero external magnetic fields[19]. The broadband spectral coverage of our THz generation and detection methods readily allows for the measurement of EPR signals in transition-metal complexes with THz-frequency ZFSs at zero field. In cases where there is more than one spin transition, the ZFS parameter absolute values can be derived from a single EPR spectrum, and the spectral linewidths are typically narrower than those at nonzero magnetic fields, yielding optimal resolution.[20] The sign of the $D$ parameter can be determined by zero-field EPR measurements at two different temperatures. The measurements can be supplemented in important ways with the addition of an applied magnetic field. Field-dependent EPR measurements allow the determination of ZFS parameters in spin systems where only one spin transition is present at zero field. Refined determination of the ZFS parameter values as well as access to the $g$-factor are also provided by field-dependent measurements. At the qualitative level, field-dependent measurements can unambiguously distinguish spin transitions from other low-frequency resonances including molecular or lattice vibrational transitions.

## EPR spectra in zero and nonzero magnetic fields

The Hamiltonian[4,21] for a single electron spin includes the ZFS and electron Zeeman interaction (EZI) terms given by,

$$\hat{H}_0 = \hat{H}_{\text{EZI}} + \hat{H}_{\text{ZFS}}. \quad (1a)$$

The ZFS Hamiltonian is commonly written as,

$$\hat{H}_{\text{ZFS}} = D\left[\hat{S}_z^2 - \frac{1}{3}S(S+1)\right] + E(\hat{S}_x^2 - \hat{S}_y^2), \quad (1b)$$

where $\hat{S}_i$ ($i = x, y, z$) are spin matrices, $S$ is the total spin quantum number, and $D$ and $E$ are the axial and transverse ZFS parameters ($E \leq D/3$).[21] Diagonalization of the ZFS Hamiltonian yields the eigenenergies and eigenvectors of the magnetic sublevels of the spin system at zero static magnetic field. With a nonzero $E$ value, the eigenstates are linear combinations of $|M_S\rangle$ states. Though $M_S$ is a "good" quantum number only when $E$ is zero, the magnetic sublevels are usually denoted by $M_S$. For integer spin systems, the degeneracy among the magnetic sublevels is completely removed by nonzero $D$ and $E$. For half-integer spin systems, degeneracy is removed among different $|M_S|$ states. Kramers degeneracy between $\pm M_S$ doublets remains, but can be removed by applying an external static magnetic field.

The static EZI term accounts for the spin system under a static magnetic field $B_0$. The EZI term is usually written as

$$\hat{H}_{EZI} = \mu_B(\hat{S}_x g_x B_{0x} + \hat{S}_y g_y B_{0y} + \hat{S}_z g_z B_{0z}), \quad (1c)$$

where $\mu_B$ is the Bohr magneton, $B_{0i}(i = x, y, z)$ is the static magnetic field, and $g_i$ $(i = x, y, z)$ is the g-factor along each molecular axis. The application of a field induces splittings and shifts of the magnetic sublevels derived from the ZFS Hamiltonian. The new eigenenergies and eigenvectors of the spin system under a static external magnetic field can be sufficiently described by $\hat{H}_{ZFS}$ and $\hat{H}_{EZI}$.

The magnetic field of electromagnetic radiation $B_1(t)$ interacts with the spin system also via the Zeeman interaction. It induces resonant magnetic dipole-allowed transitions between the magnetic sublevels. By measuring the transition frequencies, the ZFS parameters in the static spin Hamiltonian can be accessed at zero field. The Zeeman splittings and frequency shifts of the magnetic dipole-allowed transitions under a nonzero static magnetic field can also be measured, allowing determination of the g-factor.

**Experimental**

The experimental setup is a free-space THz time-domain spectroscopy system in transmission geometry as shown schematically in Fig. 1. (Details of the experimental setup are presented in the Supplementary Information Fig. S1.) The THz emitter, which was a 1-mm thick (110)-cut zinc telluride (ZnTe) crystal, was illuminated by 800-nm pulses with 100 fs duration from a commercial Ti:Sapphire amplifier at a 5-kHz repetition rate. Single-cycle THz pulses with usable bandwidths spanning from 0.1 to 2.5 THz were generated by optical rectification of the laser pulses in the ZnTe crystal.[22] The broadband, linearly polarized THz pulses were focused onto the sample where the THz magnetic fields induced magnetic dipole-allowed transitions between the magnetic sublevels.[23] The resulting spin coherences radiated electromagnetic signals, known as free-induction decays (FIDs), at the resonance frequencies. We determined the electric field profiles of the FID signals with sub-ps resolution in the time domain by measuring the THz field-induced depolarization of a variably delayed 800-nm probe pulse in another ZnTe crystal.[22,24] The sample was placed in a helium cryostat with a split superconducting magnet that could be used to apply a static magnetic field $B_0$ in the 0-5.5 T range. The orientation of $B_0$ was perpendicular to the polarization of the THz magnetic field $B_1$.

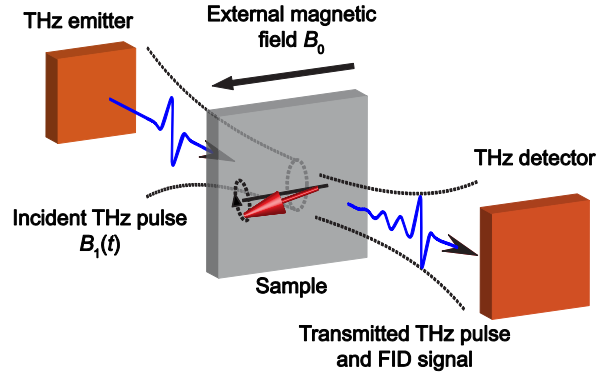

**Fig. 1.** Schematic illustration of experimental geometry. Single-cycle THz pulses are generated by illuminating the THz emitter with fs laser pulses. The THz pulses are incident onto the sample to excite the spin transitions. The transmitted THz pulses and FID signals are directed into the THz detector. Fs laser pulses are overlapped with the THz fields at the THz detector for phase-resolved detection of the THz signals. The sample is placed in a cryostat with an external magnetic field $B_0$ perpendicular to the THz magnetic field $B_1$ (e.g. the Voigt geometry shown here). Fourier transformation of the THz signals yields the EPR spectra. In this work, the THz emitter and detector are both ZnTe crystals as described in the Supplementary Information.

We derive the THz Fourier-transform (FT) amplitude spectra $|E_{sig}(\nu)|$ of the sample both at zero field and at discrete $B_0$ levels through numerical Fourier transformation of the FID signals. Absorbance spectra $A(\nu)$ of the samples are obtained by comparing the sample spectra with the reference spectra $|E_{ref}(\nu)|$ which were measured without the presence of the sample. The absorbance $A(\nu)$ in units of optical density (OD) is given by

$$A(\nu) = -2\log_{10}\left(|E_{\text{sig}}(\nu)|/|E_{\text{ref}}(\nu)|\right). \quad (2)$$

**Results and Discussion**

To demonstrate the utility of the method, we chose prototypical samples that cover all of the common transition metal high-spin (HS) states and that are of interest in a diverse spectrum of research fields in which the knowledge of ZFS parameters is crucial. The following proof-of-principle samples were chosen: hemin, a compound that is related to heme-based enzymes, containing HS Fe(III) in a square-pyramidal environment[14,17]; $CoX_2(PPh_3)_2$ (X = Cl or Br, $PPh_3$ = triphenylphosphine), a known series of compounds exhibiting single-ion single-molecule magnetic behavior stemming from HS Co(II) in a pseudo-tetrahedral coordination environment[25]; $[Fe(H_2O)_6](BF_4)_2$, a well-known integer-spin compound with HS Fe(II) in an octahedral environment[26]; and $NiCl_2(PPh_3)_2$, an integer-spin compound with HS Ni(II) in pseudo-tetrahedral rather than the usual square planar geometry[18]. Microcrystalline powders of each compound were pressed into pellets which were used in the measurements (for details see Supplementary Information).

**High-spin Fe(III): Spin-5/2 system**

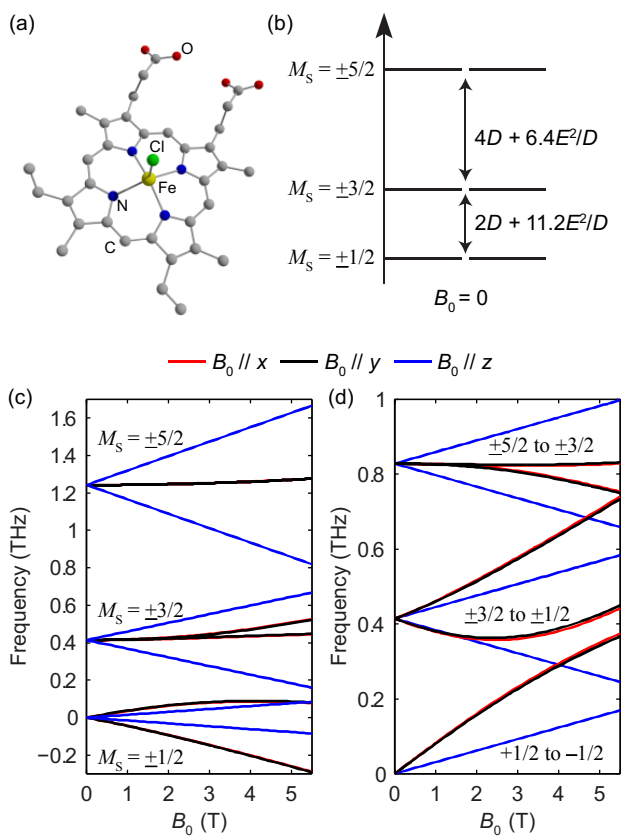

**Fig. 2.** (a) The structure of hemin. Hydrogen atoms are omitted for clarity. (b) Zero-field magnetic sublevel energy diagram of the HS Fe(III) in hemin ($S = 5/2$ spin system), where a positive $D$ value is assumed. The magnetic dipole-allowed transitions are shown by the double arrows. (c) Calculated Zeeman splitting of the magnetic sublevels as a function of $B_0$ in hemin. (d) Calculated frequencies of magnetic dipole-allowed transitions as functions of $B_0$ in hemin. In (c) and (d), the color coding indicates the direction of $B_0$ with respect to the molecular axes as shown by the legend. The black and red curves ($B_0$ // $x$ and $y$) are too closely overlapped to distinguish in these plots.

Hemin has been under extensive study in EPR, as it is related to heme, which is the functional group in heme-based metalloproteins such as hemoglobin and myoglobin. The square-pyramidal structure of hemin is shown in Fig. 2(a). The valence electrons ($d^5$) of the HS Fe(III) in hemin indicate a total spin number $S = 5/2$. The magnetic sublevels derived by diagonalizing the ZFS Hamiltonian with $S = 5/2$ are shown in Fig. 2(b), where the eigenstates are denoted by $M_S$. Magnetic dipole-allowed transitions are denoted by the double-sided arrows, and the transition frequencies shown as functions of $D$ and $E$ are calculated through second-order perturbation theory[10]. Measuring the frequencies of these two transitions at zero field readily allows determination of both the $D$ and $E$ parameter values. Zero-field measurements at two temperatures where the intermediate $M_S = \pm 3/2$ doublet states are either populated or not allows determination of the sign of $D$. The Zeeman interaction under nonzero applied field $B_0$ lifts the Kramers degeneracy and shifts the energies of the $M_S$ states as shown in Eq. (1c) and Fig. 2(c). The frequencies of the magnetic dipole-allowed transitions as functions of $B_0$ are plotted in Fig. 2(d). By measuring the frequency shifts of the transitions as a function of applied field strength, the values of the $g$-factor components can be obtained. The crystallites in the pellet sample are oriented randomly with respect to the applied magnetic field, so all three components can be determined.

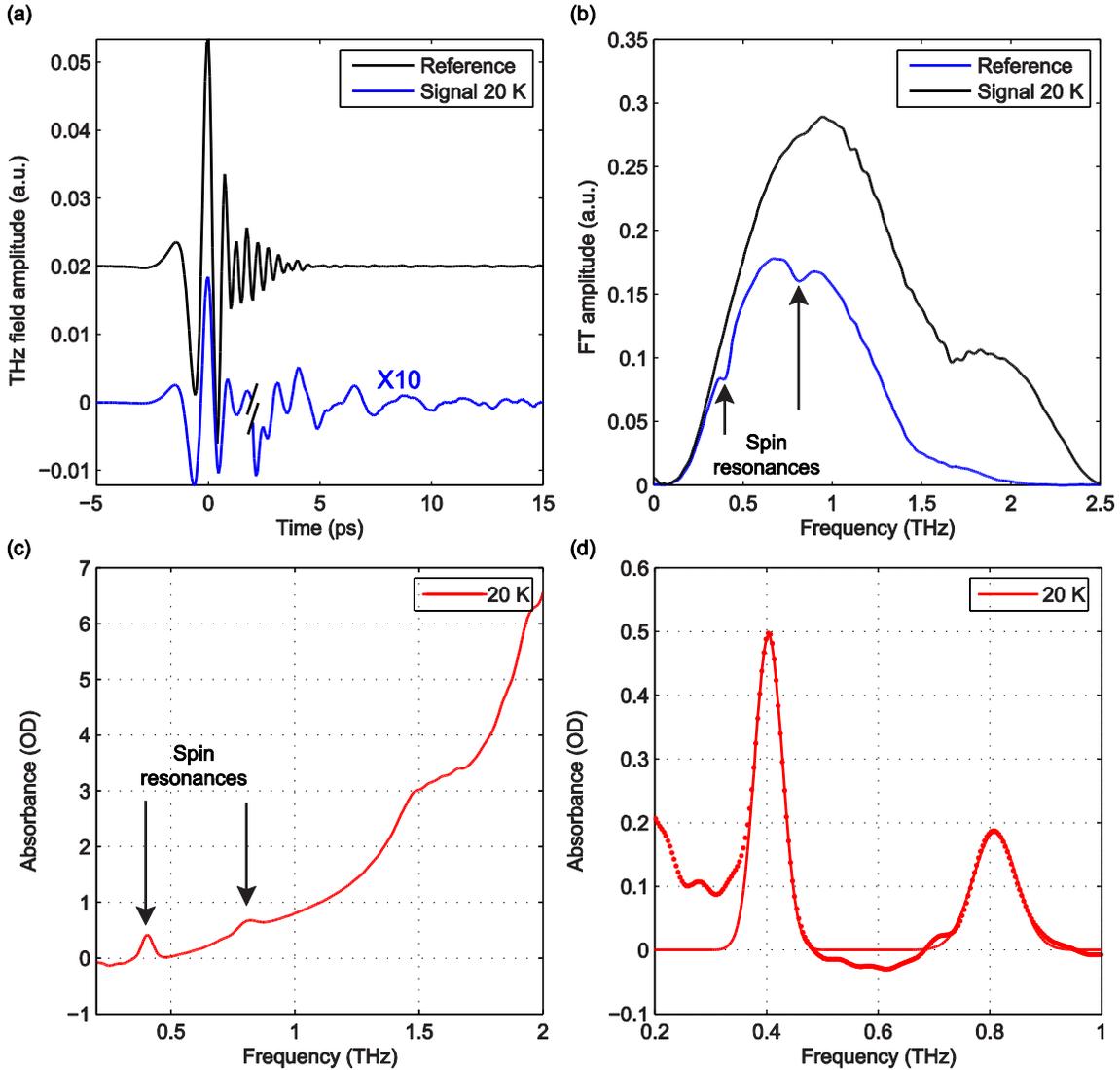

**Fig. 3.** (a) Time-domain waveform of the reference THz pulse (black) and the THz pulse transmitted through the sample followed by the FID signal (blue) at zero field. The traces are separated vertically and the FID signal is magnified by 10 for clarity. (b) FT amplitude spectra of the THz reference pulse (black) and the THz pulse and FID signal transmitted through the sample (blue). Two spin resonances are indicated by arrows. (c) Raw absorption spectrum of hemin at 20 K. (d) Absorption spectrum after background subtraction (dots) and a fit to two Gaussian functions (solid line).

The raw time-domain FID signal $E(t)$ and the FT amplitude spectrum $|E(\nu)|$ measured from hemin at zero field and 20 K are shown in Fig. 3. The time-domain waveform of the THz pulse transmitted through the sample shows attenuation and slight broadening due to the THz absorption and dispersion in the sample. The FID signal is identified as the complex waveform profile following the transmitted THz pulse shown in Fig. 3(a). Numerical Fourier transformation of the THz time-domain signals yielded the complex FT spectra of the reference and sample. The amplitude spectra are shown in Fig. 3(b), where two dips are assigned as the magnetic dipole-allowed transitions indicated in Fig. 2(b). The absorption spectrum of hemin at 20 K is plotted in Fig. 3(c). In the raw spectrum, these two peaks sit on the wing of broad higher-lying absorptions that may be due to low-frequency vibrations. The background was subtracted manually to yield Fig. 3(d). The lineshapes were fitted to two Gaussian functions, yielding the peak frequencies $0.404 \pm 0.001$ THz and $0.809 \pm 0.001$ THz. The $S = 5/2$ spin Hamiltonian was used to calculate the frequencies with variable ZFS parameters $D$ and $E$ to determine the absolute values $|D| = 6.74 \pm 0.01$ cm$^{-1}$ and $|E| = 0.048 \pm 0.048$ cm$^{-1}$ which show good agreement with previous measurements using frequency-domain FT THz EPR.[18]

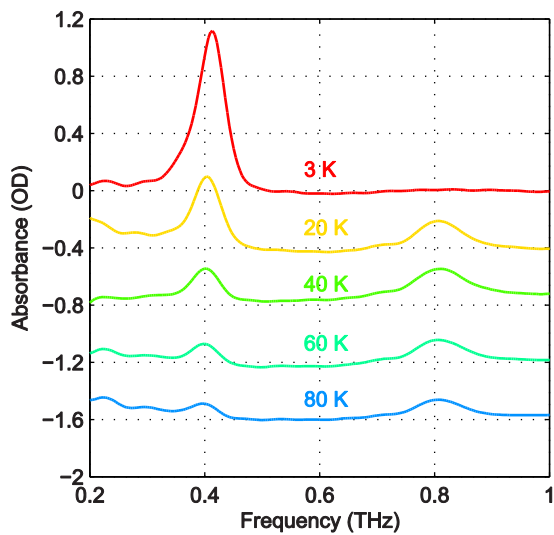

**Fig. 4.** Temperature-dependent zero-field absorbance spectra of hemin after background subtraction. The data are color-coded according to the temperatures. The higher-frequency peak disappears and the lower-frequency peak shifts slightly at reduced temperatures.

The absorbance spectra of hemin at several temperatures are shown in Fig. 4. The time-domain data and FT spectra from which the absorbance spectra were determined are shown in Fig. S2 of the Supplementary Information. At temperatures above 20 K, both peaks resulting from the two magnetic dipole-allowed transitions are present. As the temperature decreases to 3 K, the lower-frequency peak becomes more pronounced while the higher-frequency peak disappears. This shows that at low temperature the spin populations are concentrated in the $M_S = \pm 1/2$ states which must therefore be lowest in energy, indicating that the sign of $D$ is positive. If $D$ were negative, the $M_S = \pm 5/2$ states would be lowest in energy and the higher-frequency transition would predominate at low temperature. Details about determination of the sign of $D$ can be found in the Supplementary Information. The spectral peak at ~0.40 THz at 20 K shifts to ~0.41 THz at 3 K. This shift indicates a change in the ZFS parameter that may arise from subtle changes in the geometric or electronic structure of hemin at different temperatures.

Field-dependent measurements were conducted at 3 K and 20 K. $B_0$ was oriented along the THz propagation direction (i.e., Faraday geometry). The experimental absorbance spectra at zero field and six discrete $B_0$ levels are shown in Figs. 5(a) and 5(b) at 3 K and 20 K, respectively. The background was manually subtracted from all spectra. The time-domain data and FT amplitude spectra are shown in Fig. S3 of the Supplementary Information. The Kramers degeneracy is lifted and the $\pm M_S$ doublets are split at nonzero $B_0$. Due to the anisotropy in the $g$-factor, the spectral peaks are shifted by the Zeeman interaction along all three molecular axes in the powder sample used, corresponding to the field-dependent frequency shifts shown in Fig. 2(d). A new magnetic dipole-allowed transition between the $M_S = \pm 1/2$ states emerges at nonzero $B_0$ and is shifted from zero frequency to the spectral window shown as $B_0$ increases. The spectra at 20 K become somewhat crowded due to the presence of two peaks at zero field which both split to produce four peaks, plus the new peak starting from zero frequency, with some peaks merging to produce complicated lineshapes as the field strength is increased. The spectra at 3 K are simpler since only transitions originating from the $M_S = \pm 1/2$ states appear.

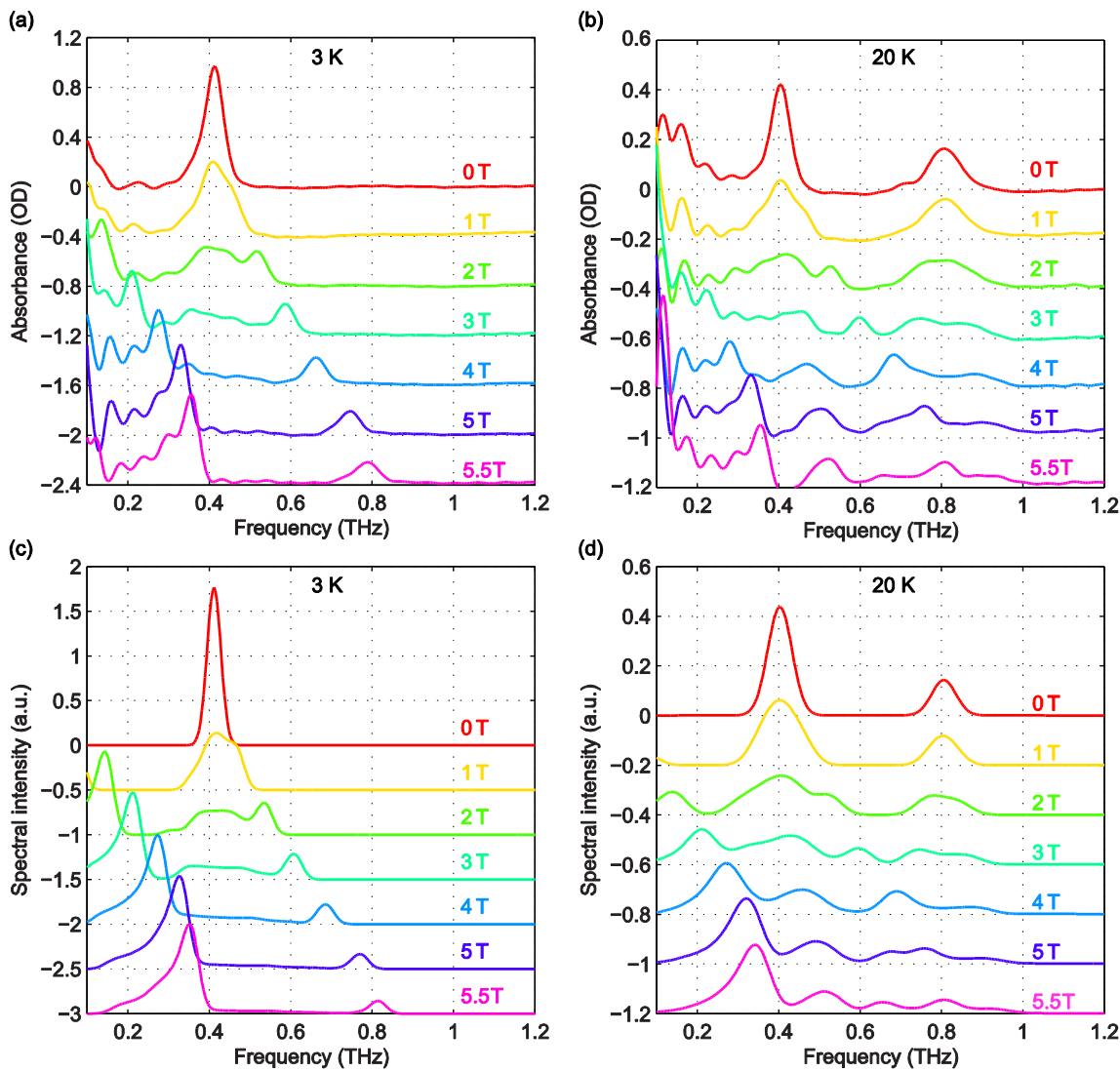

**Fig. 5.** (a) and (b) Experimental absorbance spectra of hemin as a function of $B_0$ at 3 K (a) and 20 K (b). (c) and (d) Simulated absorbance spectra at 3 K (c) and 20 K (d). The spectra are color-coded based on the values of $B_0$ indicated in the figure.

To quantitatively analyze the field-dependent spectra of our pellet samples with randomly oriented crystallites, we employed *Easyspin*, an EPR simulation software package[27]. A frequency-domain EPR simulation program was used to calculate the spectra for an $S = 5/2$ spin system at the experimental temperatures and magnetic field levels. The input parameters include the total spin quantum number (set to 5/2), the ZFS parameters $D$ and $E$, the $g$-factor elements $g_x$, $g_y$, and $g_z$, and the spectral lineshape and linewidth. The simulated spectra are shown in Fig. 5(c) and 5(d) for comparison with the experimental data. The frequency shifts and spectral lineshapes show good agreement between the experimental and simulated EPR spectra.

To eliminate possible errors introduced by the background subtraction, we plot in Figs. 6(a) and 6(b) the difference absorbance spectra between the spectra (without background subtraction) at successive $B_0$ levels.

The difference absorbance spectra can also provide enhanced sensitivity to spectral changes induced by $B_0$. The *Easyspin* simulation results at both temperatures are plotted in thick dashed lines and are overlaid with the experimental difference spectra plotted in thin solid lines in Figs. 6(a) and 6(b). The simulation results at 20 K show good agreement with the experimental data in terms of the frequency shifts and the lineshapes. The simulation results at 3 K capture the transition peak between the $M_S = \pm 1/2$ doublets due to $g_x$ and $g_y$. However, the blue-shifted spectral peak originating from the transitions between $M_S = \pm 1/2$ and $M_S = \pm 3/2$ states due to $g_x$ and $g_y$ shows slightly smaller frequency shifts than the simulated ones. This discrepancy warrants further study. Based on the comparison between the simulated and experimental field-dependent EPR spectra, we can refine the quantitative determination of the parameters in the spin Hamiltonian of hemin. The parameters of the spin Hamiltonian determined by the simulations of the difference spectra are summarized in Table 1.

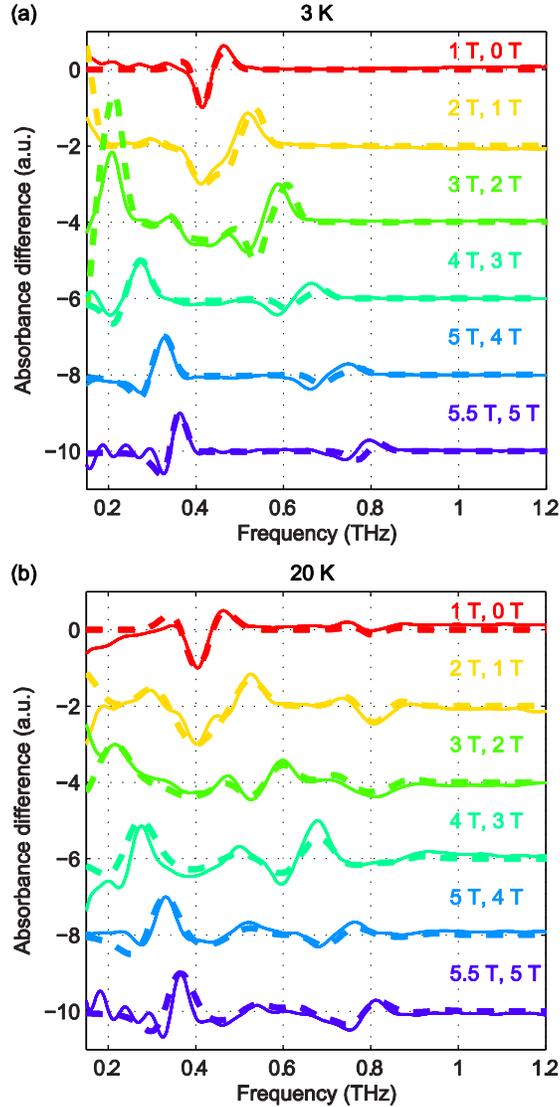

**Fig. 6.** Experimental difference absorbance spectra (solid lines) and simulated difference intensity spectra (dashed lines) at 3 K (a) and 20 K (b) at various $B_0$ levels. The spectra are color-coded according to the values of $B_0$ indicated in the figure.

**High-spin Co(II): Spin-3/2 systems**

The pseudo-tetrahedral structure of $CoX_2(PPh_3)_2$ (X = Cl or Br) is shown in Fig. 7(a).[25] The valence electrons ($d^7$) of the HS Co(II) indicate a total spin quantum number $S = 3/2$ due to three unpaired electrons.

The magnetic sublevels are derived by diagonalization of the ZFS Hamiltonian with $S = 3/2$ and are shown in Fig. 7(b) where a negative $D$ value is assumed. The eigenstates are denoted by $M_S$. Magnetic dipole-allowed transitions are denoted by the double-sided arrows, and the frequency is given by $2\sqrt{D^2 + 3E^2}$. Due to Kramers degeneracy between the $\pm M_S$ states, only one transition is observable at zero field, which is insufficient for separate determination of $D$ and $E$ or determination of the sign of $D$. The application of $B_0$ lifts the Kramers degeneracy and shifts the energies of the $M_S$ states as shown in Fig. 7(c). The field-dependent transition frequencies are calculated and plotted in Fig. 7(d). Several magnetic dipole-allowed transitions emerge under a nonzero $B_0$ and allow separate determination of $D$ and $E$. Measuring the transition between $M_S = \pm 1/2$ states at different temperatures allows determination of the sign of $D$.

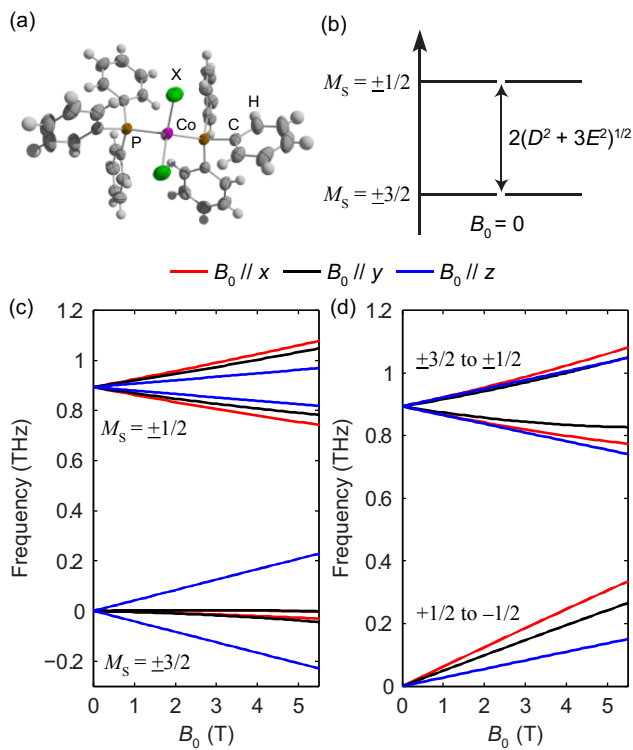

**Fig. 7.** (a) The structure of $CoX_2(PPh_3)_2$. (b) Zero-field magnetic sublevel energy diagram of the HS Co(II) in $CoX_2(PPh_3)_2$ ($S = 3/2$ spin systems), where a negative $D$ is assumed. The magnetic dipole-allowed transition is shown by the arrow. (c) Zeeman splitting of the magnetic sublevels as a function of $B_0$. (d) Frequencies of magnetic dipole-allowed transitions as functions of $B_0$. In (c) and (d), the color coding indicates the direction of $B_0$ with respect to the molecular axes as shown by the legend.

The zero-field absorbance spectra of the two compounds at low temperatures are shown in Figs. 8(a) and 8(b). The time-domain signals and FT amplitude spectra are shown in Fig. S4 of the Supplementary Information. Several strong spectral peaks are all due to vibrations. One spectral peak due to the spin transition is indicated by the arrow in each figure. The magnetic origin of the transitions is confirmed in the field-dependent measurements discussed later. Each spectral peak is fitted to a Lorenztian function, with central frequencies $0.901 \pm 0.001$ THz for $CoCl_2(PPh_3)_2$ and $0.838 \pm 0.001$ THz for $CoBr_2(PPh_3)_2$. The combination of the ZFS parameters $\sqrt{D^2 + 3E^2}$ can be determined from the peak frequency in each case, yielding $15.01 \pm 0.03$ cm$^{-1}$ and $13.96 \pm 0.03$ cm$^{-1}$ for the Cl and Br compounds respectively at their experimental temperatures. The ZFS parameter values $D = -14.76$ cm$^{-1}$ and $E = 1.141$ cm$^{-1}$ of $CoCl_2(PPh_3)_2$ have been measured by HFEPR[28], yielding $\sqrt{D^2 + 3E^2} = 14.89$ cm$^{-1}$ in good agreement with our measurement.

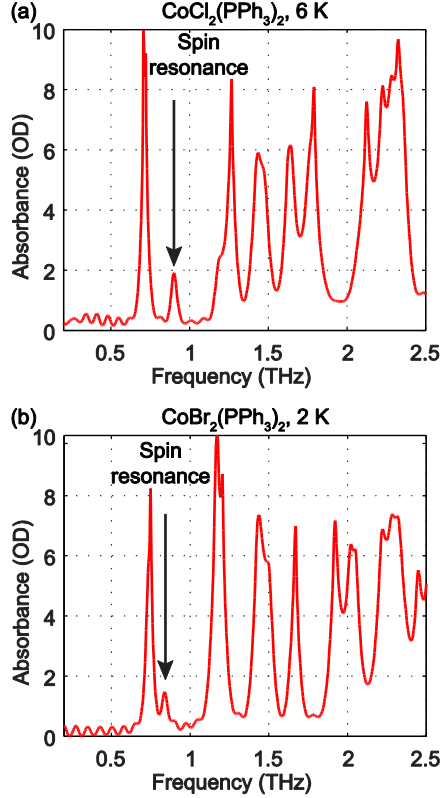

**Fig. 8.** Zero-field absorbance spectra of $CoCl_2(PPh_3)_2$ at 6 K (a) and $CoBr_2(PPh_3)_2$ at 2 K. The spin resonance peak in each figure is indicated by the arrow. Other strong absorption peaks are due to vibrations.

Field-dependent measurements from 0 to 5.5 T were conducted on each sample. $B_0$ was oriented perpendicular to the THz propagation direction (i.e., Voigt geometry). The time-domain FID data and spectra are shown in Fig. S4 of the Supplementary Information. The field-dependent absorbance spectra are shown in Fig. 9 for the two compounds. The modulations are artifacts from the Fourier transformation as discussed in the Supplementary Information. At 1 T, the spin resonance peak in each compound exhibits a decrease in amplitude and broadening due to the anisotropic Zeeman interaction. At higher $B_0$ levels, the peaks split into several peaks, following the trends shown in Fig. 7(d). The transition between the $M_S = \pm 1/2$ doublets emerges at nonzero $B_0$ and is expected to be around 0.3 THz at 5.5 T. The absence of these absorption peaks at 5.5 T at such low temperatures implies that the populations in the $M_S = \pm 1/2$ states are depleted and the populations are concentrated in the lower-energy $M_S = \pm 3/2$ states. The sign of the $D$ parameter is therefore determined to be negative in each compound.

Quantitative analysis of the field-dependent spectra is conducted by comparing the experimental difference absorbance spectra with the simulated ones. The results are shown in Fig. 10. The simulation assumes negative $D$ values for both compounds, which further confirms the determination of the sign of the $D$ parameter. Separate determination of the $D$ and $E$ parameters is possible by analyzing the field-dependent spectra for $S = 3/2$ systems. The relevant parameters of the spin Hamiltonian determined by the simulation are listed in Table 1. The measurements yield $\sqrt{D^2 + 3E^2}$ values of 15.02 cm$^{-1}$ for $CoCl_2(PPh_3)_2$ at 6 K and 14.00 cm$^{-1}$ for $CoBr_2(PPh_3)_2$ at 2 K. In these two compounds, the spin transition peaks lie between two stronger absorptions due to vibrations. The peak vibrational absorptions are so strong that even small imperfections in their subtraction result in artifacts in the difference absorbance spectra.

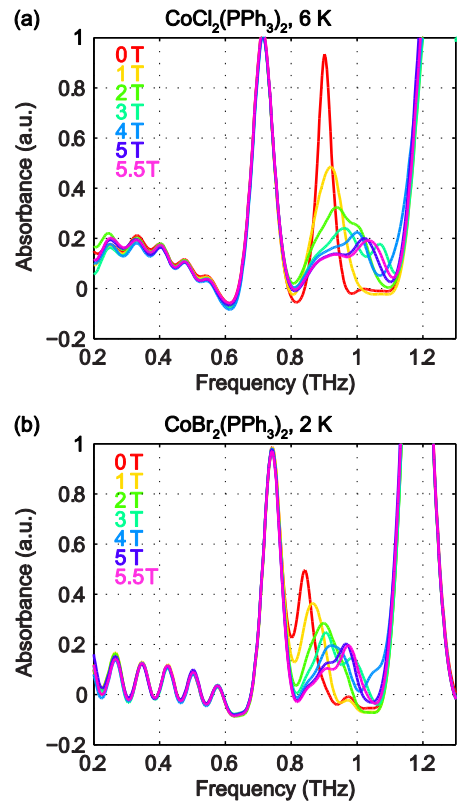

**Fig. 9.** Field-dependent absorbance spectra of $CoCl_2(PPh_3)_2$ at 6 K (a) and $CoBr_2(PPh_3)_2$ at 2 K (b). The spectral modulations are artifacts of Fourier transformation. The spectra are color-coded based on the values of $B_0$ shown in the legends.

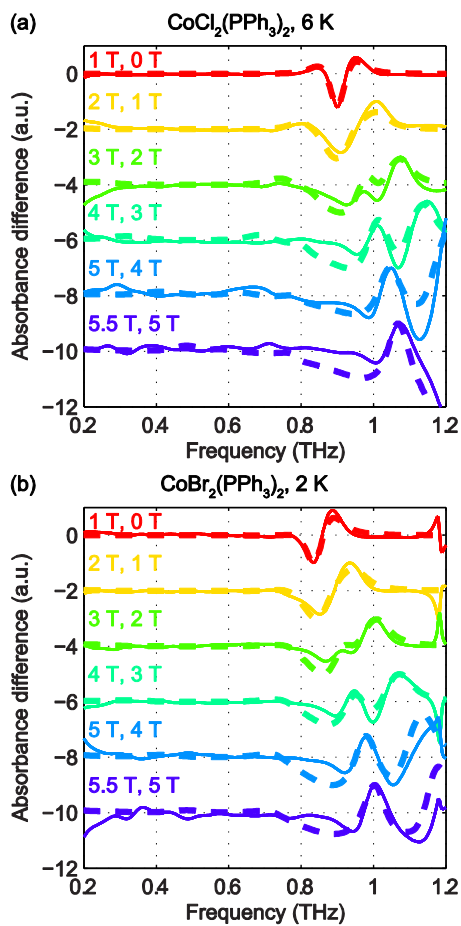

**Fig. 10.** Experimental difference absorbance spectra (solid lines) and simulated difference absorbance spectra (dashed lines) for $CoCl_2(PPh_3)_2$ at 6 K (a) and $CoBr_2(PPh_3)_2$ at 2 K (b). The spectra are color-coded according to the values of $B_0$ indicated in the figure. Additional features around 1.2 THz in (b) are due to the strong nearby vibrational absorption peak in each compound.

**High-spin Fe(II): Spin-2 system**

The structure of $Fe(H_2O)_6^{2+}$ is shown in Fig. 11(a).[29] The HS Fe(II) ion is in octahedral coordination with six water ligands. Four unpaired electrons of the valence electrons (d[6]) indicate a total spin quantum number $S = 2$. The magnetic sublevel energy diagram assuming a positive $D$ and a nonzero $E$ is shown in Fig. 11(b). The new eigenstates are denoted by $\Phi_i$ with their eigenenergies labeled in Fig. 11(b). Magnetic dipole-allowed transitions are denoted by the double-sided arrows. As the $\Phi_i$ states are superpositions of the $M_S$ states, six magnetic dipole-allowed transitions exist at zero field. As $D \gg E$ is typical, the splitting between $\Phi_4$ and $\Phi_5$ is small. The transitions to $\Phi_4$ and $\Phi_5$ states are often merged, which results in four distinct transitions. The values of the ZFS parameters $D$ and $E$ are adequately determined by a zero-field measurement of the frequencies $\nu_{12}$ and $\nu_{13}$. The sign of $D$ can be determined by temperature-dependent measurements at zero field. The application of $B_0$ further shifts the magnetic sublevels as shown in Fig. 11(c). The frequency shifts of the magnetic dipole-allowed transitions as a function of $B_0$ are shown in Fig. 11(d).

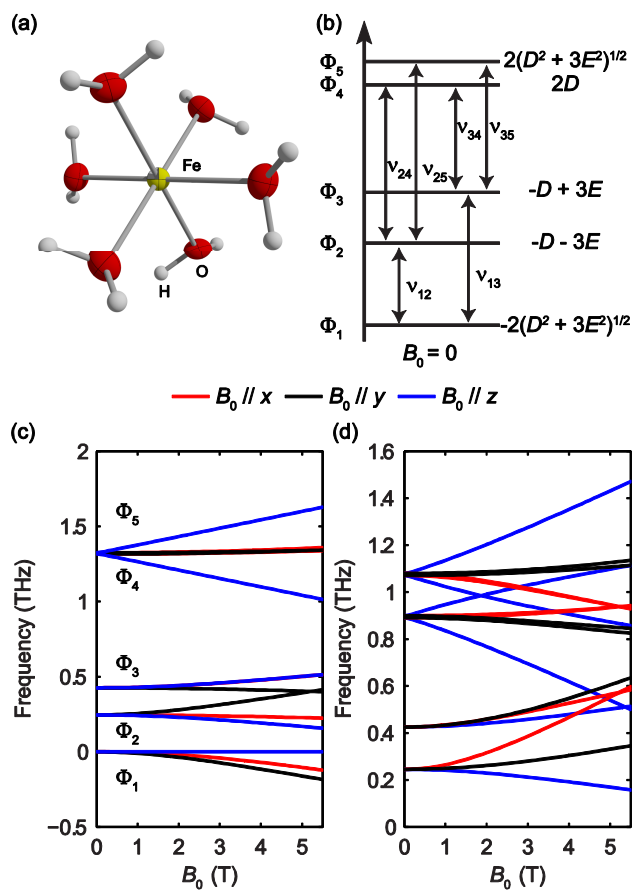

**Fig. 11.** (a) The structure of $Fe(H_2O)_6^{2+}$. (b) Zero-field magnetic sublevel diagram of the HS Fe(II) in $Fe(H_2O)_6(BF_4)_2$ ($S = 2$ spin system), where a positive $D$ value is assumed. The magnetic dipole-allowed transitions are shown by the arrows. (c) Zeeman splitting of the magnetic sublevels as a function of $B_0$. (d) Frequencies of magnetic dipole-allowed transitions as functions of the external magnetic field. In (c) and (d), the color coding indicates the direction of $B_0$ with respect to the molecular axes, as shown by the legend.

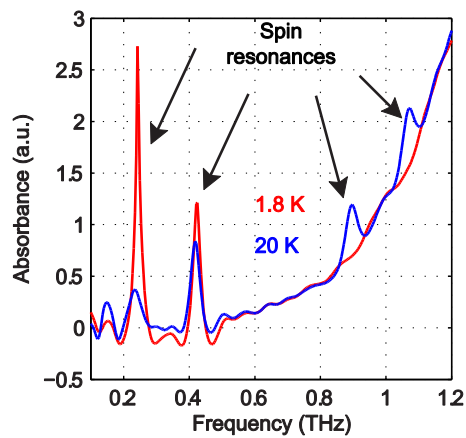

**Fig. 12.** Zero-field absorbance spectra of $Fe(H_2O)_6(BF_4)_2$ at 1.8 K (red) and 20 K (blue). At 1.8 K, two strong peaks are assigned as the spin resonances resulting from two transitions. At 20 K, four peaks are assigned as the spin resonances resulting from six transitions. In this case the peak frequencies could be determined accurately without subtraction of the background absorption.

The zero-field absorbance spectra of $Fe(H_2O)_6(BF_4)_2$ at 1.8 K and 20 K are shown in Fig. 12. The time-domain signals and FT amplitude spectra are shown in Fig. S5 of the Supplementary Information. At 1.8 K, only the lowest state $\Phi_1$ is populated. Two absorption peaks located through Lorentzian fits at $0.243 \pm 0.001$ THz and $0.423 \pm 0.001$ THz are assigned as the spin transitions at frequencies $\nu_{12}$ and $\nu_{13}$ between $\Phi_1$ and $\Phi_2$ and between $\Phi_1$ and $\Phi_3$, respectively, as shown in Fig. 11(b). The assignments were confirmed by field-dependent measurements discussed below. Based on the values of $\nu_{12}$ and $\nu_{13}$, the ZFS parameters are calculated to be $|D| = 10.82 \pm 0.03$ cm$^{-1}$ and $|E| = 1.00 \pm 0.01$ cm$^{-1}$. At 20 K, four transitions are observed as the intermediate states $\Phi_2$ and $\Phi_3$ are also populated. The two absorption peaks at frequencies $\nu_{12}$ and $\nu_{13}$ become weaker, and two additional spectral peaks at ~0.90 THz and ~1.07 THz emerge due to the transitions between the higher-lying states. Hence the sign of the $D$ parameter is determined to be positive.

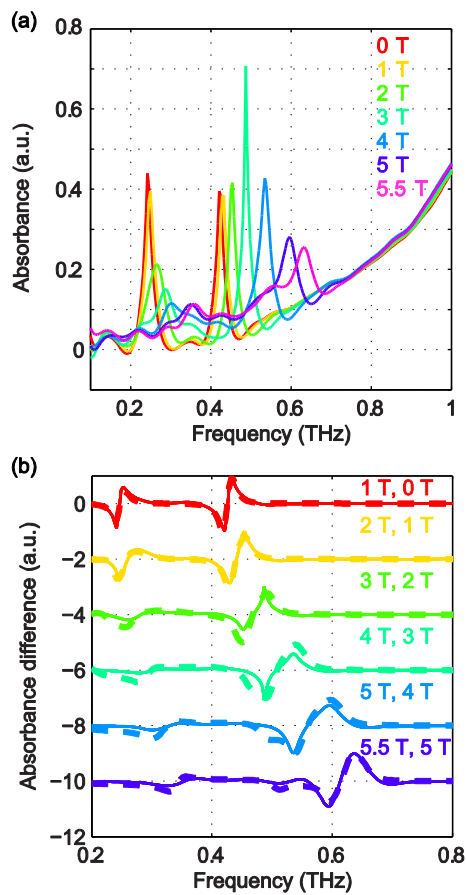

**Fig. 13.** (a) Field-dependent absorbance spectra of Fe(H$_2$O)$_6$(BF$_4$)$_2$ at 1.8 K. The two spin resonance peaks show splittings and shifts as a function of the applied magnetic field. (b) Experimental difference absorbance spectra (solid lines) and simulated difference intensity spectra (dashed lines) for Fe(H$_2$O)$_6$(BF$_4$)$_2$ at 1.8 K. The spectra are color-coded according to the magnetic field values indicated in each figure.

Due to the relative simplicity of the zero-field EPR spectra at 1.8 K where only two transitions appear, field-dependent measurements from 0 to 5.5 T were conducted at 1.8 K in the Voigt geometry. The experimental absorbance spectra are shown in Fig. 13(a). The time-domain signals and FT amplitude spectra are shown in Fig. S6 of the Supplementary Information. The spectral peaks assigned as spin resonances show splittings and shifts as a function of $B_0$, which confirm their magnetic origins. To eliminate the background due to the wing of higher-lying absorptions by vibrations, difference absorbance spectra as a function of $B_0$ are shown in Fig. 13(b). The spectra are overlapped with simulated ones for comparison, which show excellent agreement. The spin Hamiltonian parameters determined by the simulations are listed in Table 1.

**High-spin Ni(II): Spin-1 system**

The pseudo-tetrahedral structure of NiCl$_2$(PPh$_3$)$_2$, which is similar to the structure of CoCl$_2$(PPh$_3$)$_2$, is shown in Fig. 14(a). Two unpaired electrons of the valence electrons (d$^8$) of the HS Ni(II) indicate a total spin quantum number $S = 1$. Similar to the case of the $S = 2$ spin system, with a zero $E$ value, the degeneracy between the $\pm M_S$ states remains and the eigenstates are the $M_S$ states. With a nonzero $E$ value, the degeneracy between the $M_S = \pm 1$ states is lifted and the new eigenstates are superpositions of the $M_S$ states. The magnetic sublevel energy diagram with positive $D$ and nonzero $E$ values is shown in Fig. 14(b) with new eigenstates denoted by $\Phi_i$ whose eigenenergies are indicated. Magnetic dipole-allowed transitions are denoted by the double-sided arrows. At zero field, two transitions are observable with frequencies denoted as $\nu_{12}$ and $\nu_{13}$. The values of the ZFS parameters $D$ and $E$ are adequately determined by measuring $\nu_{12}$ and $\nu_{13}$. The sign of $D$ can be determined by temperature-dependent measurements at zero field. The application of $B_0$ shifts the magnetic sublevels as shown in Fig. 14(c). The frequency shifts of the magnetic dipole-allowed transitions as a function of $B_0$ are shown in Fig. 14(d) which allows the determination of the $g$-factor.

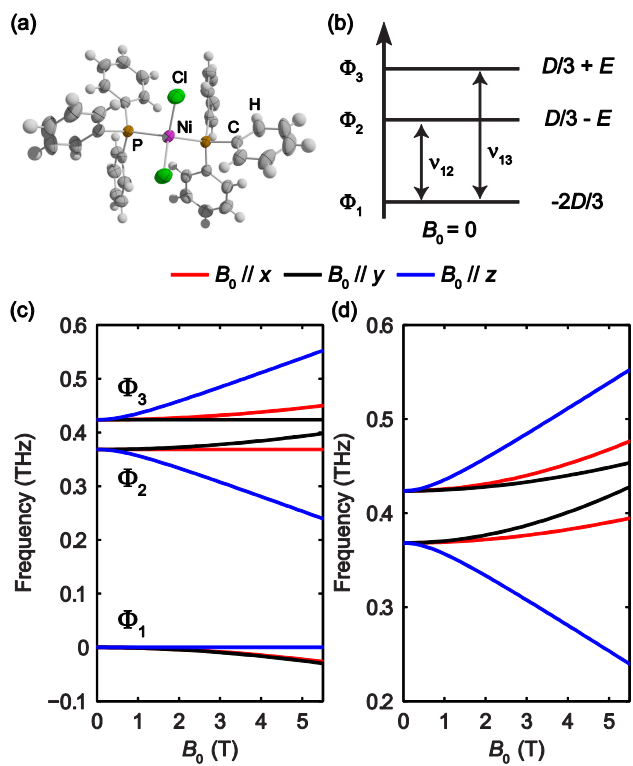

**Fig. 14.** (a) The structure of NiCl$_2$(PPh$_3$)$_2$. (b) Zero-field magnetic sublevel diagram of the HS Fe(II) in Fe(H$_2$O)$_6$(BF$_4$)$_2$ ($S = 2$ spin system), where a positive $D$ value is assumed. The magnetic dipole-allowed transitions are shown by the arrows. (c) Zeeman splitting of the magnetic sublevels as a function of $B_0$. (d) Frequencies of magnetic dipole-allowed transitions as functions of $B_0$. In (c) and (d), the color coding indicates the direction of $B_0$ with respect to the molecular axes, as shown by the legend.

The zero-field absorbance spectra of NiCl$_2$(PPh$_3$)$_2$ at 2 K and 10 K are shown in Fig. 15. The time-domain signals and FT amplitude spectra are shown in Fig. S7 of the Supplementary Information. The dominant peak is likely due to a lattice vibrational mode. It is at a frequency similar to that of a peak in the spectra of CoCl$_2$(PPh$_3$)$_2$ due to their similar molecular structures and atomic masses. Two weaker absorption peaks are assigned as the spin resonances. The ZFS parameters can be calculated based on the frequencies of these two peaks, which correspond to the transitions at $\nu_{12}$ and $\nu_{13}$ in Fig. 14(b). The frequencies were obtained by fitting the spectral lineshapes to two Gaussian functions, yielding $\nu_{12} = 0.337 \pm 0.001$ THz and $\nu_{13} = 0.459 \pm 0.001$ THz. The ZFS parameters were calculated to be $|D| = 13.27 \pm 0.03$ cm$^{-1}$ and $|E| = 2.03 \pm 0.03$ cm$^{-1}$ at 2 K, which show good agreement with previous HFEPR results[28] yielding $D = 13.196$ cm$^{-1}$ and $E = 1.848$ cm$^{-1}$. The appearance of two peaks at both 2 K and 10 K implies that the lowest energy state is $\Phi_1$, where the population is concentrated. If $\Phi_3$ were the lowest energy state, the population at $\Phi_2$ would be mostly depleted and only one transition peak would be expected at 2 K. Hence the sign of the $D$ parameter is determined to be positive.

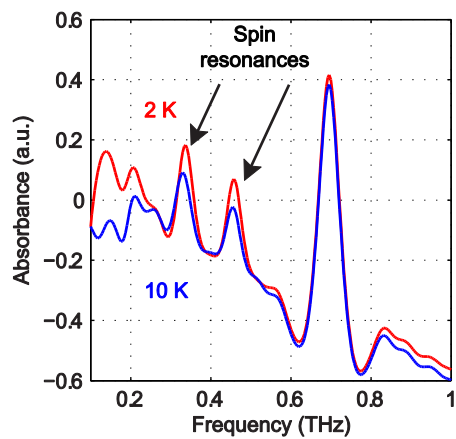

**Fig. 15.** Zero-field absorbance spectra of NiCl$_2$(PPh$_3$)$_2$ at 2 K (red) and 10 K (blue). The two spin resonance peaks are shown by the arrows and are present at both temperatures. The stronger peak is due to a vibration.

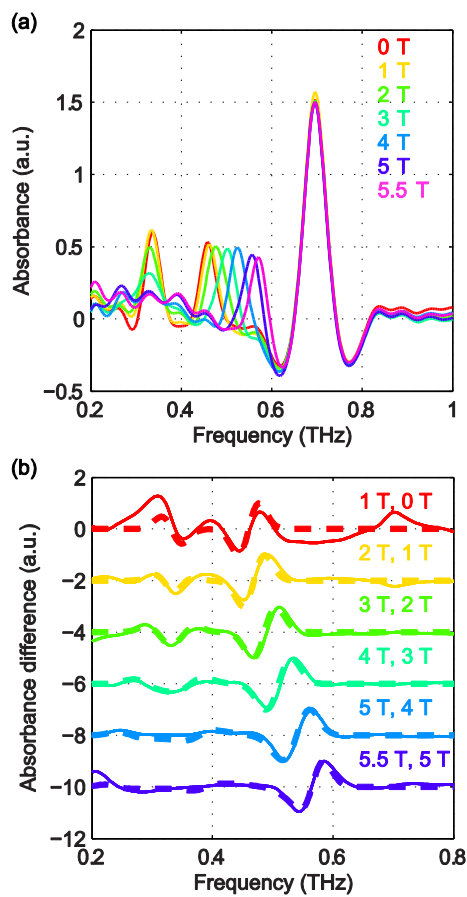

**Fig. 16.** (a) Field-dependent absorbance spectra of NiCl$_2$(PPh$_3$)$_2$ at 2 K. The two spin resonance peaks show splittings and shifts as a function of $B_0$. The strong absorption peak is likely due to lattice vibrations. (b) Experimental difference absorbance spectra (solid lines) and simulated difference intensity spectra (dashed lines) for NiCl$_2$(PPh$_3$)$_2$ at 2 K. The spectra are color-coded according to the values of $B_0$ indicated in each figure.

The field-dependent absorbance spectra of NiCl$_2$(PPh$_3$)$_2$ at 2 K (Voigt geometry) are shown in Fig. 16(a). The time-domain signals and FT amplitude spectra are shown in Fig. S8 of the Supplementary Information. The two spin resonance peaks show field-dependent frequency shifts. The dependence of the spectral peaks on the magnetic field is analyzed by comparing the difference absorbance spectra with those obtained from simulations. The results are shown in Fig. 16(b), which shows excellent agreement between the experimental and simulated spectra. The spin Hamiltonian parameters determined by the simulations are listed in Table 1.

**Table 1.** Spin Hamiltonian parameters determined through simulations of the field-dependent difference EPR spectra for the compounds.

|  | $D$ (cm$^{-1}$) | $E$ (cm$^{-1}$) | $g_x$ | $g_y$ | $g_z$ |
|---|---|---|---|---|---|
| Hemin, 3 K | 6.90 | 0 | 1.91 | 1.91 | 2.05 |
| Hemin, 20 K | 6.73 | 0.02 | 1.91 | 1.91 | 2.05 |
| CoCl$_2$(PPh$_3$)$_2$, 6 K | –14.76 | 1.61 | 2.20 | 2.18 | 2.23 |
| CoBr$_2$(PPh$_3$)$_2$, 2 K | –13.9 | 0.96 | 2.10 | 2.10 | 2.22 |
| Fe(H$_2$O)$_6$(BF$_4$)$_2$, 1.8 K | 10.82 | 1.00 | 2.10 | 2.10 | 2.10 |
| NiCl$_2$(PPh$_3$)$_2$, 2 K | 13.27 | 2.00 | 2.20 | 2.17 | 2.17 |

## Conclusions

We demonstrate using four representative molecules a fast, facile, and reliable technique to measure ZFS parameters directly, in the absence of a magnetic field, and to refine the determination through field-dependent measurements. We use THz time-domain FID measurements to obtain EPR signals associated with THz-frequency ZFSs in molecular complexes. We fully characterized the values and signs of the ZFS parameters for several compounds belonging to $S = 1, S = 3/2, S = 2$ and $S = 5/2$ spin systems based on the zero-field and/or field-dependent EPR spectra. Values of the $g$-factor are also obtained from the field-dependent measurements. This technique permits unambiguous assignment of THz-frequency ZFS parameters at different temperatures, which is difficult to accomplish by magnetometry measurements. More specifically, integer THz-frequency spin systems were termed "EPR silent" as the magnetic dipole-allowed transitions are not accessible in traditional EPR measurements due to the large ZFS parameters. HFEPR can access complexes with moderate ZFSs that are shifted into the excitation bandwidth via the Zeeman interaction, but the magnetic fields required for such measurements with monochromatic excitation sources can be quite large for complexes with high ZFS. The THz time-domain EPR measurement provides a direct way to measure large ZFSs in the THz-frequency region which is characteristic for molecular magnets.

In our current experiments, the spin number density was on the order of $10^{21}$ cm$^{-3}$. Considering the ~5-mm THz beam diameter throughout the 2-mm thick pellet samples, the measured signals emerged from ~$10^{20}$ spins (see Supplementary Information for details). The estimated sensitivity of our current measurement configuration is $10^{19}$ spins which is close to that required from a reasonable amount of large biomolecules[15,17]. In this work, the THz source and detector utilized allow us to measure resonances between about 6 and 80 cm$^{-1}$. The ranges of $D$ and $E$ that are accessible in this work depend on the specific spin system under study. Although we used homebuilt THz systems for our measurements, existing THz technologies[30] including some commercially available tabletop instruments can provide broader THz spectral coverage throughout the far-infrared range[31], higher resolution[32], higher sensitivity[33–36], and faster data acquisition times[32,37]. These will allow determination of ZFS parameters and other magnetic fine structure revealed through THz-frequency spin transitions.

Because the technique is independent of the identity of the spin center and of the spin value, it can be applied broadly to systems with non-zero ZFS. We expect that the general approach in using THz time-domain spectroscopy to characterize transitions in the spin manifold of open-shell systems can be further elaborated to include other magnetic interactions, including magnetic exchange, for instance. Apart from the demonstrated advantages of THz time-domain spectroscopy over its frequency-domain counterpart,[38] the time-domain oscillation periods of ~1 ps will allow measurement of the dynamic evolution of unpaired electron spins as revealed by time-dependent changes in the ZFS spectra following, for example, pulsed optical excitation of a molecular state from which charge transfer or spin crossover occurs[39,40]. Two-dimensional (2D) THz magnetic resonance spectroscopy of collective spin waves (magnons) has recently been demonstrated[41], and 2D THz EPR measurements of HS compounds may also prove possible. THz EPR echoes (as observed in the 2D THz measurements of magnons) could prove useful for separation of spin transitions in biomolecules from low-frequency vibrations which may undergo very rapid dephasing, after which echo signals may be dominated by otherwise obscured spin coherences. The methodology presented herein and its future developments will find wide-ranging applications in characterizing magnetic properties of molecules, biological systems, and condensed matter.

**Acknowledgements**

We thank Stephen Hill for stimulating discussions and Christopher Hendon for the artwork. The spectroscopic work by JL, XL, BKO-O, and KAN was supported by Office of Naval Research Grant No. N00014-13-1-0509 and Defense University Research Instrumentation Program Grant No. N00014-15-1-2879, National Science Foundation Grant No. CHE-1111557, and the Samsung Global Research Outreach program. Measurements done in magnetic field by OO, CB and NG were supported by US Department of Energy, BES DMSE, Award number DE-FG02-08ER46521. CB acknowledges support from the National Science Foundation Graduate Research Fellowship under Grant No. 1122374. Synthetic work by GS, LS, and MD was supported by the Center for Excitonics, an Energy Frontier Research Center funded by the Basic Energy Sciences program of the US Department of Energy Office of Science (award No. DESC0001088 to the Massachusetts Institute of Technology [MIT]).

**Notes and references**

# Rapid and Precise Determination of Zero-Field Splittings by Terahertz Time-Domain Electron Paramagnetic Resonance Spectroscopy

Jian Lu[1], I. Ozge Ozel[2], Carina Belvin[2], Xian Li[1], Grigorii Skorupskii[1], Lei Sun[1], Benjamin K. Ofori-Okai[1], Mircea Dincă[1], Nuh Gedik[2], and Keith A. Nelson[1,*]

[1]*Department of Chemistry, Massachusetts Institute of Technology, Cambridge, Massachusetts 02139, USA*

[2]*Department of Physics, Massachusetts Institute of Technology, Cambridge, Massachusetts 02139, USA*

\* kanelson@mit.edu

**Supplementary Information**

# Contents

# 1. Experimental setup

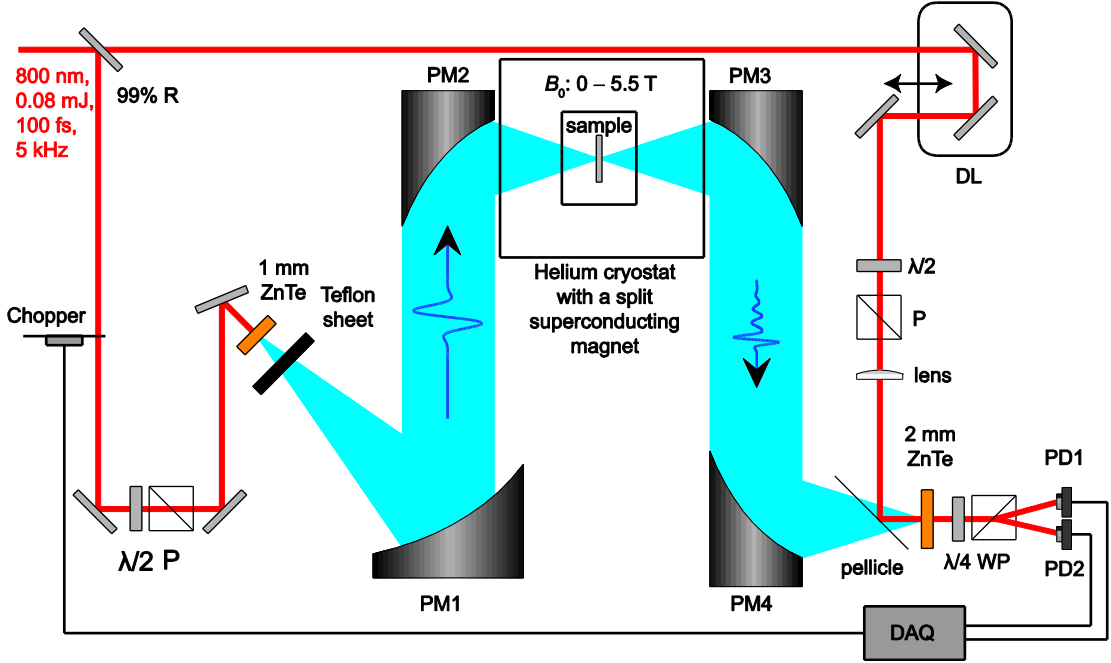

**Figure S1.** THz time-domain spectroscopy system used in the experiments. DL: delay line; ZnTe: zinc telluride; PM: parabolic mirror; λ/2: half wave-plate; P: polarizer; λ/4: quarter wave-plate; WP: Wollaston prism; PD: photodiode; DAQ: data acquisition card. Red lines indicate optical paths and black lines indicate electrical connections. The blue shaded areas indicate THz beam paths.

Our experimental setup for the tabletop THz time-domain EPR spectroscopy system is shown in basic form in Fig. 1 of the main paper and in more detail schematically in Fig. S1. The laser was a commercial Ti:sapphire amplifier (Spitfire Pro, Spectra Physics) delivering 800 nm pulses with duration of 100 femtoseconds at a repetition rate of 5 kHz. The total output power of 400 mW was split into two optical paths by a 92/8 pellicle beamsplitter. The stronger pulses were modulated at 2.5 kHz by an optical chopper and were incident onto a 1-mm (110)-cut ZnTe crystal to generate THz pulses via optical rectification[1–3]. Single-cycle THz pulses were generated from the ZnTe crystal with a useable bandwidth spanning from 0.1 to 2.5 THz. The residual laser light transmitting through the ZnTe crystal was blocked by a black Teflon sheet. The THz pulses were collimated by a 45-degree off-axis parabolic mirror (PM) and focused onto the sample by a 90-degree off-axis PM (PM1 and PM2, respectively, in Fig. S1). The THz pulses transmitted through the sample and the FID signals that followed them were collimated and focused into a 2-mm ZnTe detection crystal by a pair of 90-degree off-axis PMs (PM3 and PM4 in Fig. S1). The weaker laser pulses from the beamsplitter were time delayed by a delay line (a mechanical translation stage) and attenuated by a half wave-plate (λ/2) and a polarizer (P). They were subsequently focused and overlapped with the THz beam in the ZnTe detection crystal to measure the phase-resolved THz signals via electro-optic sampling[4]. In this measurement, THz electric fields induced a modulation of the refractive indices of the ZnTe crystal along two orthogonal directions. The laser pulses experienced a transient birefringence due to this modulation. The THz-induced birefringence was measured as intensity modulations of the two optical polarization components which were separated by a quarter wave-plate (λ/4) and a Wollaston prism (WP) and detected by a pair of photodiodes (PD1 and PD2). The difference between the measured intensities was detected by a data acquisition card (DAQ) triggered by the chopper. The temporal profiles of the THz pulses and the FID signals were measured with sub-picosecond time resolution by scanning the delay line. The THz beam path was kept under dry air purge, which suppressed THz absorption due to water vapor in the atmosphere. The dynamic range of spectral amplitude of the system was in excess of $10^3$ (corresponding to a dynamic range of spectral intensity in excess of $10^6$). Further discussions of the underlying principles of THz generation and detection by optical rectification and electro-optic sampling can be found in References 1-3. The samples were placed in a helium cryostat with a split superconducting magnet (SuperOptiMag, Janis) which could provide static magnetic fields $B_0$ ranging from 0 to 5.5 T. The orientation of $B_0$ was perpendicular to the polarization of the THz magnetic field $B_1$. $B_0$ can be either parallel or perpendicular to the propagation direction of the THz pulse. The former geometry is called Faraday geometry and the latter is called Voigt geometry[5]. These two geometries do not result in any differences in the EPR measurements. All the elements used in the setup are commercially available.

The time-domain signals measured experimentally typically had 390 time steps of 67 fs. At each time point, the signal was typically averaged for 1000 laser shots. Under these conditions, the data acquisition time for a single absorbance spectrum was roughly 4 minutes, including measuring the time-domain signals for both the reference and the sample. The absorbance spectra reported below were averaged data from 10 measurements, which took roughly 40 minutes to collect. The time window of 26 ps was limited by the THz double reflection in the 1-mm ZnTe crystal used for THz generation. The resulting instrument-limited frequency resolution was approximately 39 GHz. Each spectrum reported here was interpolated by zero padding of the time-domain signals to 4096 data points in the data processing.

In some cases, the oscillatory FID signals emerging from the samples last longer than the instrument-limited time window of 26 ps. The spectra show ringing artifacts due to Fourier transformation of the raw time-domain signals with square windows. A Hamming apodization function was applied to the absorbance spectra, which reduced the ringing effects and had minimal effects on the frequencies of the spin resonance signals of interest.

## 2. Time-domain waveforms and FT amplitude spectra

### 2.1 Hemin

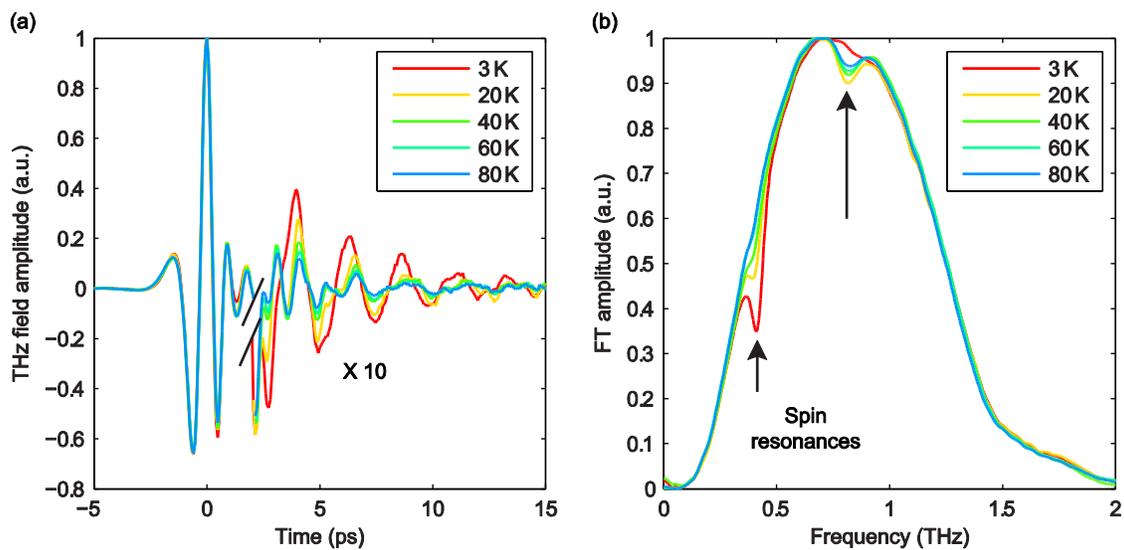

**Figure S2.** (a) Time-domain waveform of THz pulses transmitted through a pellet of hemin at various temperatures. Oscillatory features following the transmitted THz pulse are identified as the FID signals which are magnified by 10. (b) FT amplitude spectra of the THz pulses transmitted through hemin at various temperatures obtained by numerical Fourier transformation of time-domain waveforms in (a). The spin resonances are indicated by the arrows. The data are color-coded according to the temperatures shown in the legends.

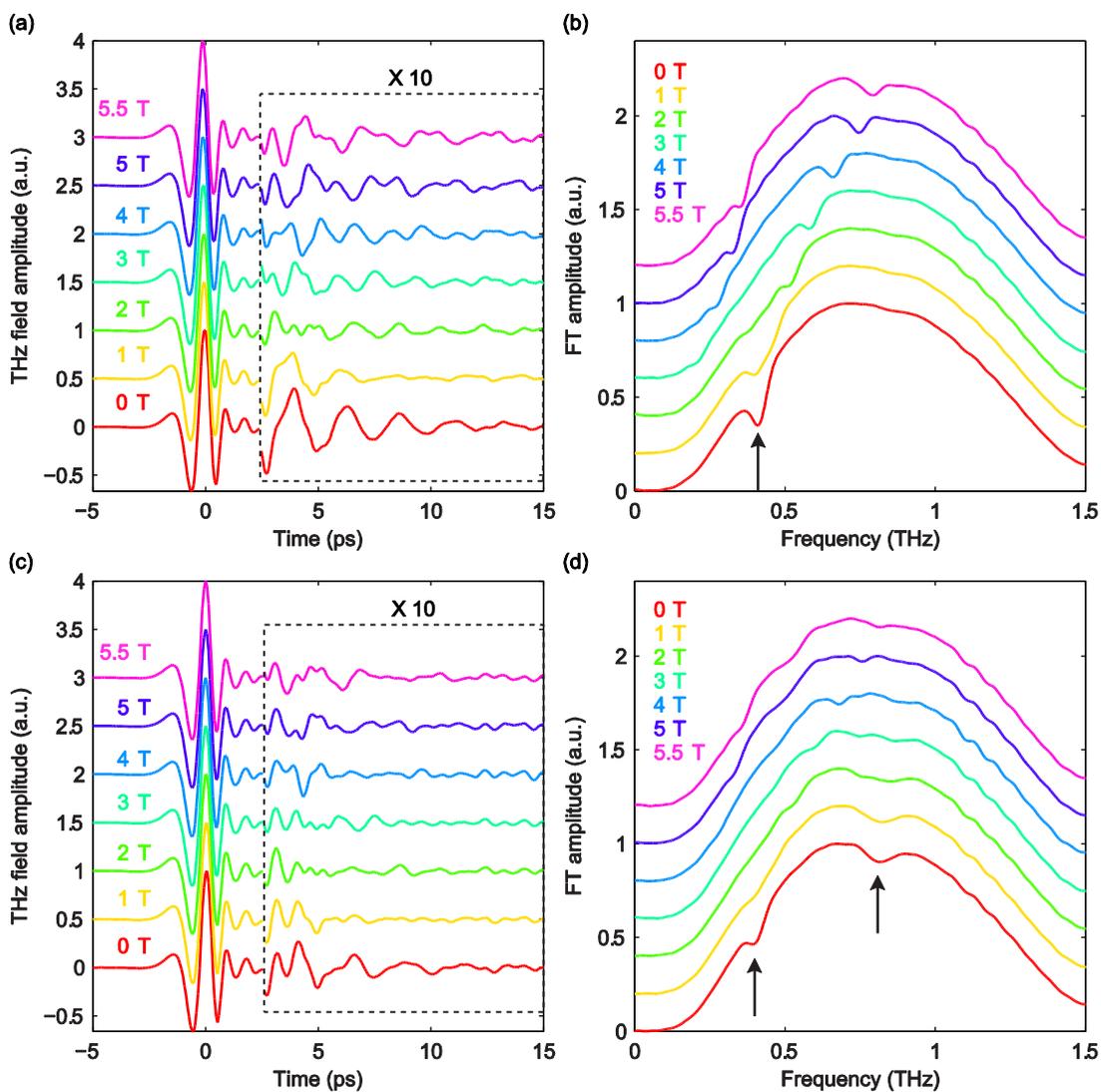

**Figure S3.** (a) Time-domain FID traces and (b) FT amplitude spectra as functions of $B_0$ for hemin at 3 K. (c) Time-domain FID traces, and (d) FT amplitude spectra as functions of $B_0$ for hemin at 20 K. The time-domain traces and spectra are color-coded according to the values of $B_0$ indicated. In (a) and (c), the FID signals in the dashed boxes are magnified by 10 to bring out the weak signals. In (c) and (d), the arrows indicate the spin resonances at zero field.

## 2.2 CoX$_2$(PPh$_3$)$_2$

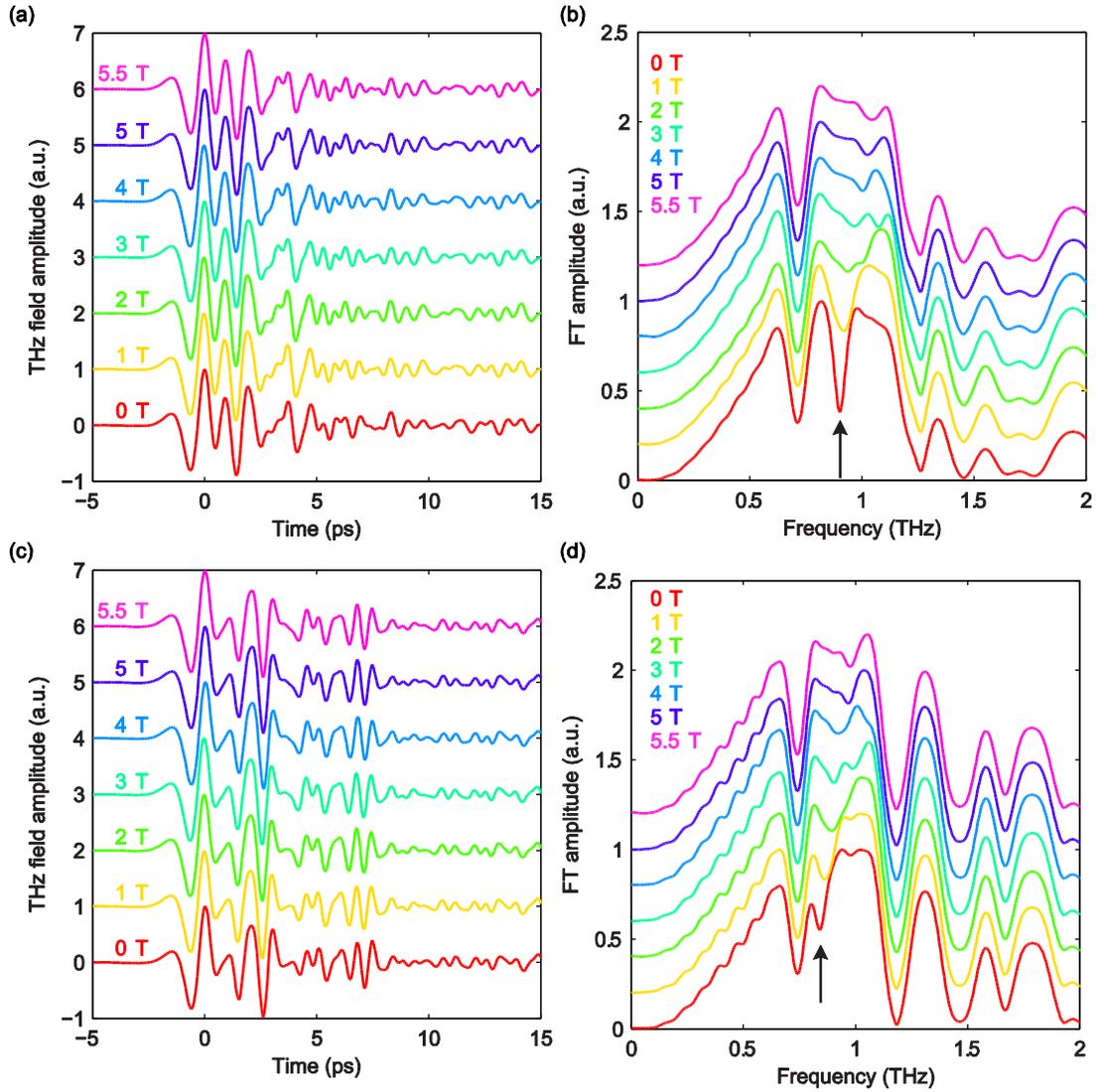

**Figure S4.** (a) Time-domain FID traces and (b) FT amplitude spectra as functions of $B_0$ for CoCl$_2$(PPh$_3$)$_2$ at 6 K. (c) Time-domain FID traces, and (d) FT amplitude spectra as functions of $B_0$ for CoCl$_2$(PPh$_3$)$_2$ at 2 K. The time-domain traces and spectra are color-coded according to the values of $B_0$ indicated. In (a) and (c), the signals are dominated by vibrational FIDs. In (c) and (d), a Hamming apodization function was applied to each spectrum. The arrows indicate the spin resonances at zero field.

## 2.3 Fe(H$_2$O)$_6$(BF$_4$)$_2$

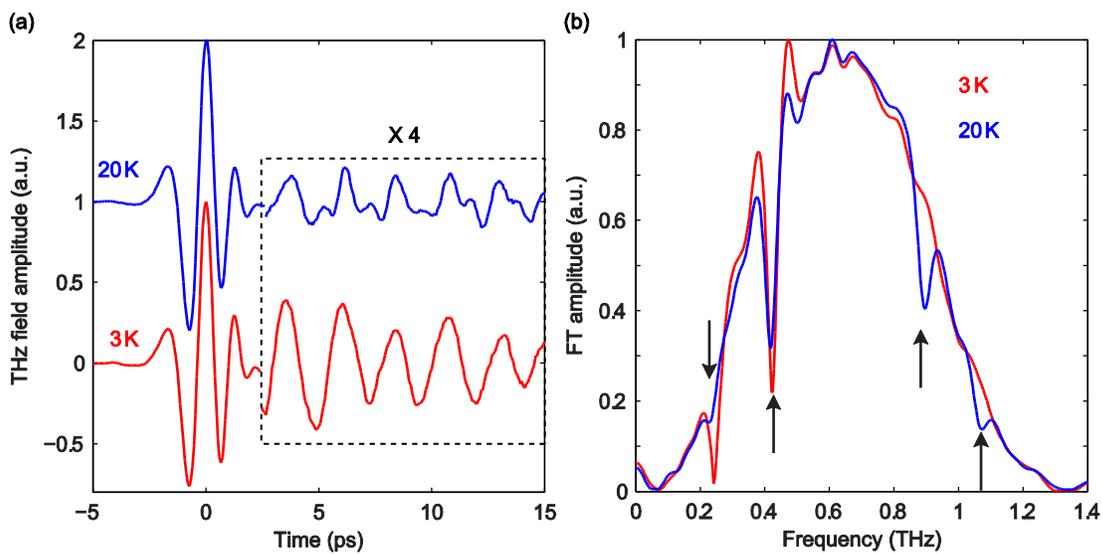

**Figure S5.** (a) Time-domain FID traces and (b) FT amplitude spectra for Fe(H$_2$O)$_6$(BF$_4$)$_2$ at 1.8 K (red) and 20 K (blue). The traces and spectra are color-coded according to the temperatures indicated. In (a), the FID signals are magnified by 4. In (b), a Hamming apodization function was applied to each spectrum. The arrows indicate the spin resonances at zero field.

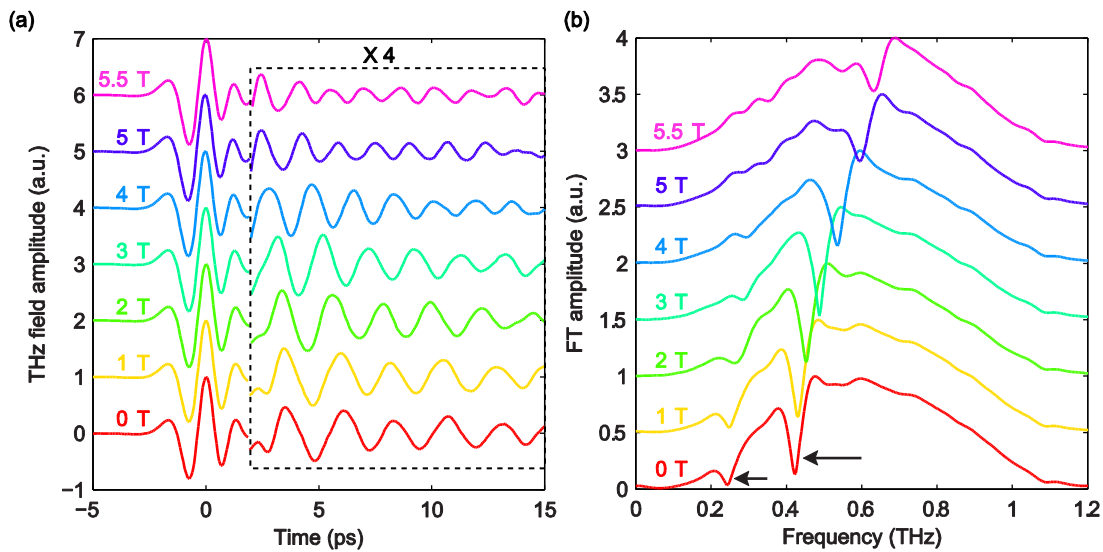

**Figure S6.** (a) Time-domain FID traces and (b) FT amplitude spectra as functions of $B_0$ for Fe(H$_2$O)$_6$(BF$_4$)$_2$ at 1.8 K. The traces and spectra are color-coded according to the values of $B_0$ indicated. In (a), the FID signals in the dashed box are magnified by 4. In (b), a Hamming apodization function was applied to each spectrum.

## 2.4 NiCl$_2$(PPh$_3$)$_2$

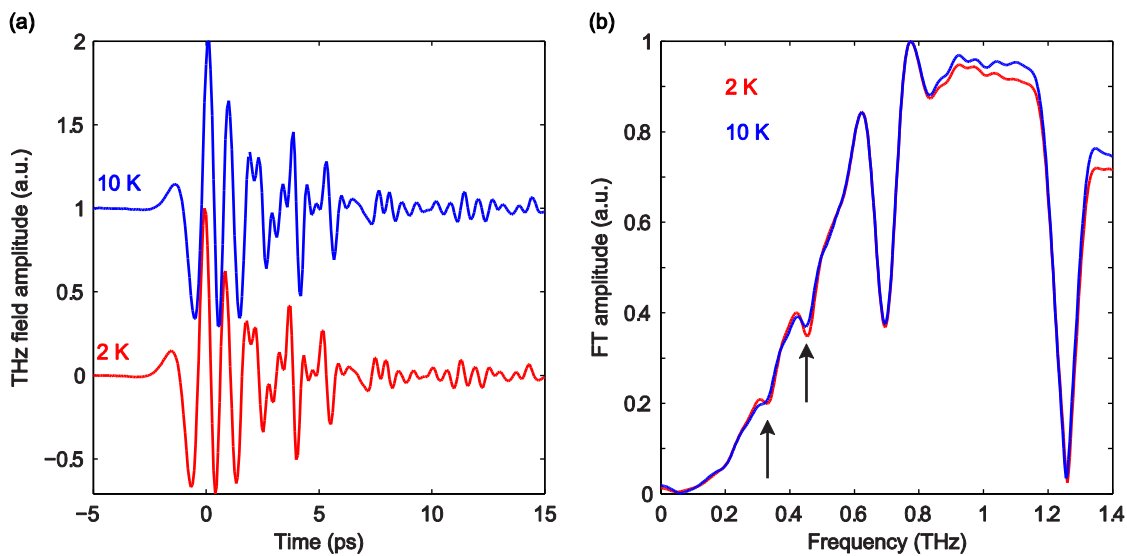

**Figure S7.** (a) Time-domain FID traces and (b) FT amplitude spectra for NiCl$_2$(PPh$_3$)$_2$ at 2 K (red) and 10 K (blue). The traces and spectra are color-coded according to temperatures indicated. In (a), the signals are dominated by vibrational FIDs. In (b), a Hamming apodization function was applied to each spectrum. The arrows indicate the spin resonances at zero field.

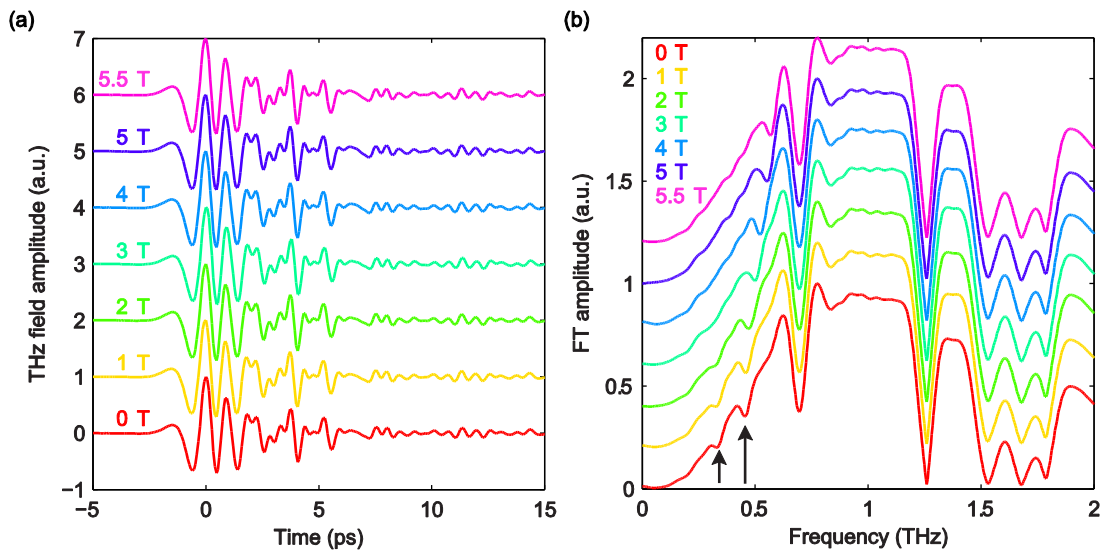

**Figure S8.** (a) Time-domain FID traces and (b) FT amplitude spectra as functions of $B_0$ for NiCl$_2$(PPh$_3$)$_2$ at 2 K. The time-domain traces and spectra are color-coded according to the values of $B_0$ indicated. In (a), the FID signals are dominated by vibrational signals. In (b), a Hamming apodization function was applied to each spectrum. The arrows indicate the spin resonances at zero field.

## 3. Spin Hamiltonian

The spin Hamiltonian[6,7] discussed in the main text consists of the ZFS and EZI terms. The general form of the ZFS parameter $\bar{\bar{D}}$ is a second-rank tensor, which is set traceless (the sum of the diagonal components is zero) and symmetric ($D_{ij} = D_{ji}$). The general form of the spin Hamiltonian describing the ZFS for a single spin system is written as

$$\hat{H}_{ZFS} = \hat{\mathbf{S}} \cdot \bar{\bar{D}} \cdot \hat{\mathbf{S}}_0^T, \quad (S1)$$

where $\hat{\mathbf{S}} = [\hat{S}_x \quad \hat{S}_y \quad \hat{S}_z]$ is the spin vector, and $\hat{S}_i$ ($i = x, y, z$) are the spin matrices. In the eigenframe where the $D$ tensor is diagonal, the ZFS Hamiltonian can be transformed into the commonly written form as,

$$\hat{H}_{ZFS} = D\left[\hat{S}_z^2 - \frac{1}{3}S(S+1)\right] + E(\hat{S}_x^2 - \hat{S}_y^2), \quad (S2)$$

where $S$ is the total spin quantum number, and $D$ and $E$ are the axial and transverse ZFS parameters, given by[7] $D = \frac{3}{2}D_{zz}$ and $E = \frac{D_{xx} - D_{yy}}{2}$. For a spin-$S$ system, the spin matrices are square matrices of dimension $2S + 1$. They are always represented in the Zeeman basis with states $|S, M_S\rangle$ ($M_S = -S, -S+1, \ldots, S$). The states are usually denoted by $|M_S\rangle$ for short. The spin matrices can be constructed using the following relations

$$\langle M_S' | \hat{S}_x | M_S \rangle = \left(\delta_{M_S', M_S+1} + \delta_{M_S'+1, M_S}\right)\frac{1}{2}\sqrt{S(S+1) - M_S' M_S}, \quad (S3a)$$

$$\langle M_S' | \hat{S}_y | M_S \rangle = \left(\delta_{M_S', M_S+1} - \delta_{M_S'+1, M_S}\right)\frac{1}{2i}\sqrt{S(S+1) - M_S' M_S}, \quad (S3b)$$

$$\langle M_S' | \hat{S}_x | M_S \rangle = \delta_{M_S', M_S} M_S, \quad (S3c)$$

where $\delta$ is the Kronecker delta. The ZFS energy levels of the spin systems studied in the main text can be derived from these equations.

The general form of the EZI term is written in the tensor form as

$$\hat{H}_{EZI} = \mu_B \hat{\mathbf{S}} \cdot \bar{\bar{g}} \cdot \mathbf{B}_0^T, \quad (S4)$$

where $\mu_B$ is Bohr magneton and $\mathbf{B}_0 = [B_{0x} \quad B_{0y} \quad B_{0z}]$ is the applied static magnetic field vector, and $\bar{\bar{g}}$ is the g-factor, which is a tensor. The g-factor is usually symmetric and can be transformed into diagonal form, with the diagonal elements $g_i$ ($i = x, y, z$) determined from experimental measurements.

## 4. Determination of $D$ and $E$ parameters from zero-field and field-dependent EPR measurements

As shown in the main paper, zero-field EPR measurements yield the absolute values $|D|$ and $|E|$ of the ZFS parameters for integer spin systems, in which the spin sublevels are nondegenerate. For $S = 3/2$ systems, degeneracies among the sublevels limit the zero-field EPR measurements to determination of combinations of the ZFS parameters, as $D$ and $E$ cannot be separately determined from one doubly degenerate transition at $\sqrt{D^2 + 3E^2}$ measured in zero-field EPR. Application of an external magnetic field shifts the levels, enabling separate determination of the absolute values $|D|$ and $|E|$ for $S = 3/2$ systems. For other half-integer spin systems, zero-field EPR can measure more than one (doubly degenerate) magnetic dipole-allowed transition derived from the magnetic sublevels, so $|D|$ and $|E|$ can be determined without application of an external field.

Variation of the temperature in the presence of a magnetic field allows determination of the sign of $D$ for $S = 3/2$ systems and for $S = 1$ systems with zero $E$ parameter values. In all other spin systems, variation of the temperature at zero magnetic field is sufficient for determination of the sign of $D$. The details with examples of $S = 3/2$, $S = 5/2$, $S = 1$ and $S = 2$ systems are elaborated briefly below.

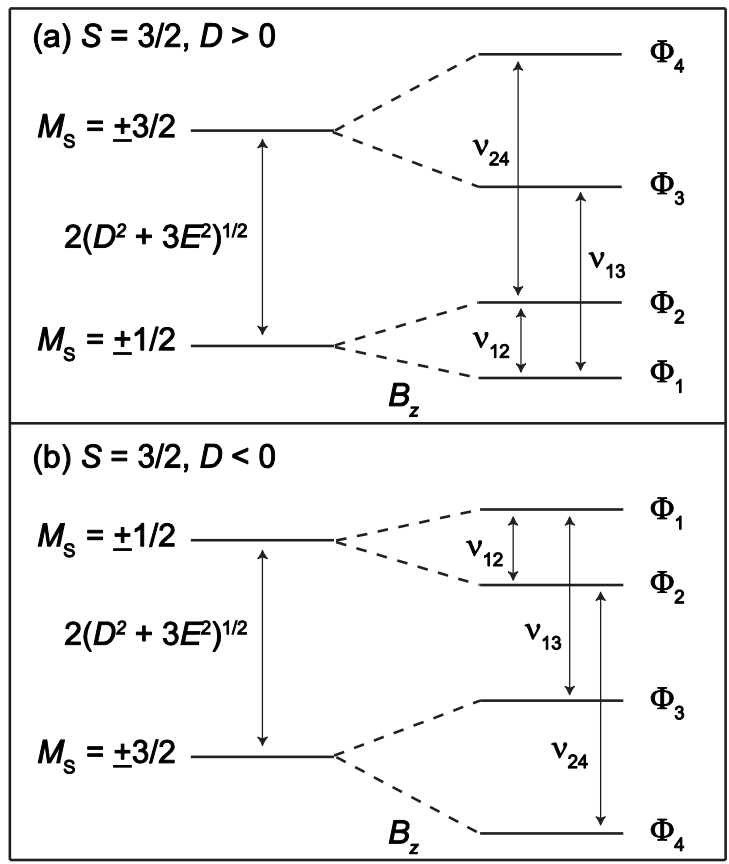

**Figure S9.** Magnetic sublevel energy diagrams for $S = 3/2$ systems with a positive $D$ (a) or a negative $D$ (b). Though the magnetic sublevels are labeled with $M_S$, note that $M_S$ is a "good" quantum number only for $E = 0$. In each case, applying an external magnetic field along the molecular $z$-axis, the Kramers doublets are split which allows determination of the sign of $D$.

In order to determine the sign of the $D$ parameter in $S = 3/2$ systems, EPR spectra with varying external magnetic field and temperature are necessary. As an example, we show in Fig. S9 the magnetic sublevel energy diagrams for $S = 3/2$ systems with positive or negative $D$. Without splitting the Kramers doublets, the EPR transition frequencies are both $\sqrt{D^2 + 3E^2}$. We assume an external magnetic field $B_z$ along the molecular $z$-axis, which can be achieved with a single-crystal sample. (For powder samples, the molecular orientation is random with respect to the magnetic field. Zeeman interactions with an anisotropic $g$ tensor result in somewhat complex lineshapes and need to be analyzed numerically. ZFS and $g$ tensor parameters can be obtained with high precision from field-swept EPR measurements combined with spin Hamiltonian simulations[5,8].) The applied field splits the Kramers doublets, separating the two transitions that are overlapped in the zero-field spectrum and also introducing a new magnetic dipole-allowed transition at frequency $\nu_{12}$ between the $\Phi_1$ and $\Phi_2$ levels (derived from $M_S = \pm 1/2$ states that were degenerate under zero field). The spectral amplitude of the new peak exhibits different trends for different signs of $D$ as the temperature is varied, due to the Boltzmann factor. For positive $D$, states $\Phi_1$ and $\Phi_2$ have lower energies than states $\Phi_3$ and $\Phi_4$ and the $\nu_{12}$ transition has an increasing spectral amplitude as temperature is reduced and population is increased in $\Phi_1$. For negative $D$, states $\Phi_1$ and $\Phi_2$ have the higher energies and the $\nu_{12}$ transition has a decreasing spectral amplitude as temperature decreases. The external magnetic field also allows separate determination of $D$ and $E$. The frequencies of the magnetic dipole-allowed transitions shown in Fig. S9 are analytically given by

$$\nu_{12} = g_z \mu_B B_z + (D^2 + 3E^2 - 2D g_z \mu_B B_z + g_z^2 \mu_B^2 B_z^2)^{\frac{1}{2}} - (D^2 + 3E^2 + 2D g_z \mu_B B_z + g_z^2 \mu_B^2 B_z^2)^{\frac{1}{2}}; \quad \text{(S5a)}$$

$$\nu_{13} = 2(D^2 + 3E^2 - 2D g_z \mu_B B_z + g_z^2 \mu_B^2 B_z^2)^{\frac{1}{2}}; \quad \text{(S5b)}$$

$$\nu_{24} = 2(D^2 + 3E^2 + 2D g_z \mu_B B_z + g_z^2 \mu_B^2 B_z^2)^{\frac{1}{2}}. \quad \text{(S5c)}$$

By analyzing the field-dependences of these three transition frequencies from field-swept EPR spectra, we can obtain the $g_z$ factor and the $D$ and $E$ parameters.

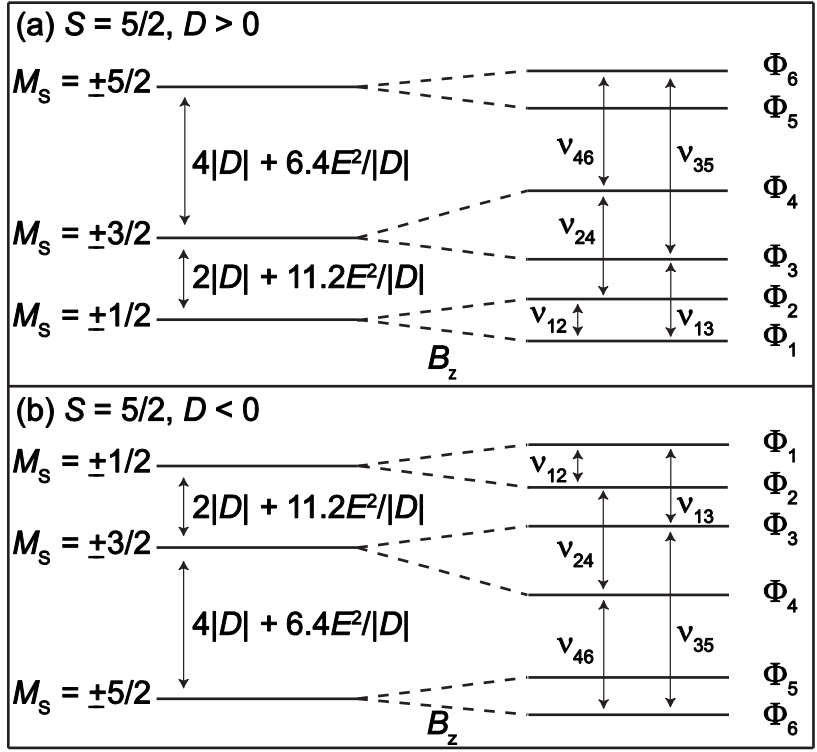

**Figure S10.** Magnetic sublevel energy diagrams for $S = 5/2$ systems with (a) positive or (b) negative $D$. Though the magnetic sublevels are labeled with $M_S$, note that $M_S$ is a "good" quantum number only for $E = 0$. The transition frequencies are calculated through second-order perturbation theory. $D$ and $E$ can both be determined from the two transition frequencies measured at zero field. The sign of $D$ can be determined through temperature-dependent measurements at zero field.

In half-integer spin systems other than $S = 3/2$, the degeneracies of magnetic dipole-allowed transitions are partially removed by nonzero $D$, which allows separate determination of $|D|$ and $|E|$ and determination of the sign of $D$ at zero field. The magnetic sublevel energy diagrams for $S = 5/2$ systems are shown in Fig. S10. Zero-field EPR measurements at a fixed temperature measure the magnetic dipole-allowed transitions between $M_S = \pm 1/2$ and $M_S = \pm 3/2$ and between $M_S = \pm 3/2$ and $M_S = \pm 5/2$, whose frequencies are calculated by second-order perturbation theory[9] to be $2|D| + 11.2\, E^2/|D|$ and $4|D| + 11.2\, E^2/|D|$ in both cases. Hence for spin-5/2 systems, $D$ and $E$ can be separately determined by measuring these two frequencies at zero external magnetic

field. The sign of $D$ can be deduced from zero-field EPR spectra with varying temperature. If $D$ is positive, $M_S = \pm 1/2$ states are lowest in energy. As temperature decreases, the lower-frequency transition amplitudes between $M_S = \pm 1/2$ and $M_S = \pm 3/2$ states increase monotonically while the higher-frequency transition amplitude between $M_S = \pm 3/2$ and $M_S = \pm 5/2$ states increases first and then decreases as the thermal population in states $M_S = \pm 3/2$ decreases. If $D$ is negative, $M_S = \pm 5/2$ states are lowest in energy. As temperature decreases, the higher-frequency transition amplitude between $M_S = \pm 5/2$ and $M_S = \pm 3/2$ states increases monotonically while the lower-frequency transition amplitude between $M_S = \pm 3/2$ and $M_S = \pm 1/2$ states increases first and then decreases as the thermal population in states $M_S = \pm 3/2$ decreases. Applying an external magnetic field further splits the degenerate doublets, allowing determination of the $g$-factor through field-dependent EPR measurements.

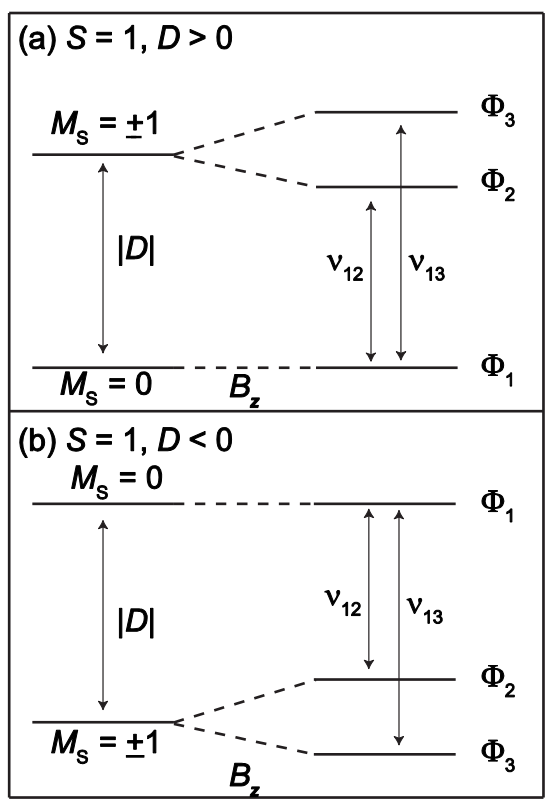

**Figure S11.** Magnetic sublevel energy diagrams for $S = 1$ systems with $E = 0$ and (a) positive or (b) negative $D$. Applying an external magnetic field $B_z$ along the molecular $z$-axis, the $M_S = \pm 1$ doublet is split, which allows determination of the sign of $D$.

For $S = 1$ systems with a zero (or nonzero) $E$ parameter, zero-field EPR measurement yields $|D|$ in both cases. An external magnetic field is required to split the doublet with $M_S = \pm 1$ as shown in Fig. S11. The sign of $D$ can then be determined by the temperature-dependent changes in spectral amplitudes of the magnetic dipole-allowed transitions. As temperature decreases, the spectral amplitudes at $\nu_{12}$ and $\nu_{13}$ both increase monotonically if $D$ is positive. If $D$ is negative, the higher-frequency transition amplitude at $\nu_{13}$ increases monotonically while the lower-frequency transition amplitude at $\nu_{12}$ increases first and then decreases as the thermal population in state $\Phi_2$ decreases. Nonzero $E$ values also can be determined through zero-field EPR measurements as illustrated in the main paper.

The magnetic sublevel energy diagrams for $S = 2$ systems are shown in Fig. S12. With $E = 0$, zero-field EPR measurements at a fixed temperature measure the magnetic dipole-allowed transitions between $M_S = 0$ and $M_S = \pm 1$ and between $M_S = \pm 1$ and $M_S = \pm 2$, whose frequencies are $|D|$ and $3|D|$ respectively regardless of the sign of $D$. As temperature decreases, the lower-frequency transition amplitudes between $M_S = 0$ and $M_S = \pm 1$ states increase monotonically while the higher-frequency transition amplitude between $M_S = \pm 1$ and $M_S = \pm 2$ states increases first and then decreases as the thermal population in states $M_S = \pm 1$ decreases if $D$ is positive. If $D$ is negative, as temperature decreases, the higher-frequency transition amplitude between $M_S = \pm 1$ and $M_S = \pm 2$ states increases monotonically while the lower-frequency transition amplitude between $M_S = 0$ and $M_S = \pm 1$ states increases first and then decreases as the thermal population in states $M_S = \pm 1$ decreases. With nonzero $E$, all the degeneracy is removed as shown in the main paper. Once again, based on different trends of the spectral amplitudes as functions of temperature one can determine the sign of $D$.

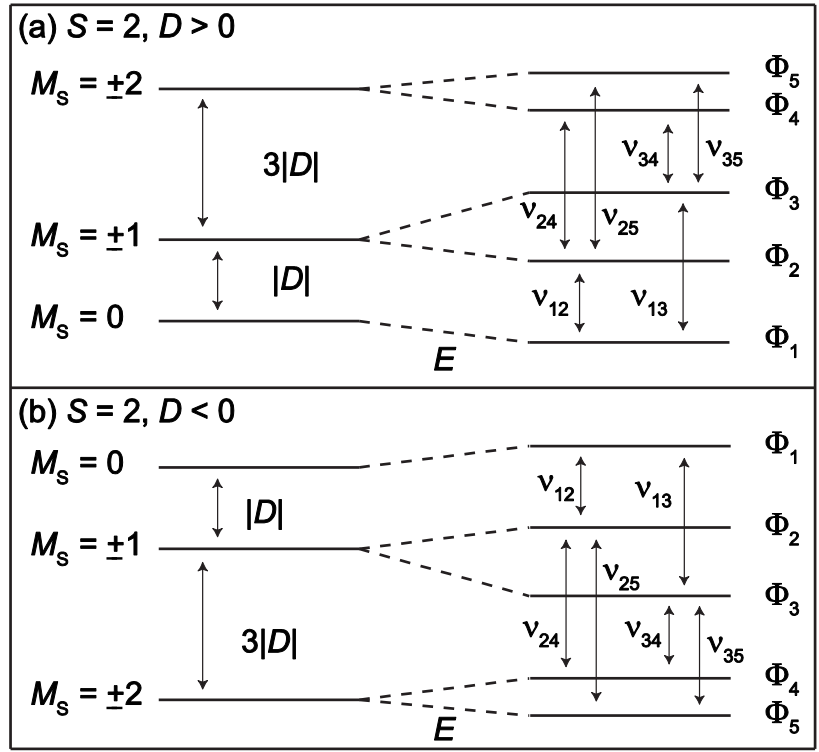

**Figure S12.** Magnetic sublevel energy diagrams for $S = 2$ systems with (a) positive or (b) negative $D$. With either zero $E$ or nonzero $E$, the sign of $D$ can be determined from the temperature-dependent spectral amplitudes.

## 5. Materials and pellet preparation

Microcrystalline powders of the two Co(II) compounds were synthesized according to literature procedures[10]. Anhydrous $CoCl_2$ and $CoBr_2$ were prepared by heating $CoCl_2 \cdot 6H_2O$ (Alfa Aesar) and $CoBr_2 \cdot xH_2O$ (Alfa Aesar), respectively, to 300 °C under dynamic vacuum (100 mtorr). Triphenylphosphine (Alfa Aesar), and absolute ethanol (VWR) were used as received. Microcrystalline powders of hemin, $[Fe(H_2O)_6](BF_4)_2$, and $NiCl_2(PPh_3)_2$ were purchased from a commercial source (Sigma Aldrich) and were used without further purification.

Approximately 200 mg of hemin and $CoX_2(PPh_3)_2$ were pressed into pellets of 13-mm diameter and ~2 mm thickness. Approximately 200 mg of $Fe(H_2O)_6(BF_4)_2$ and $NiCl_2(PPh_3)_2$ were mixed with 100 mg high-density polyethylene and the solid solutions were pressed into pellets of 13-mm diameter and ~2 mm thickness. The measurements were conducted on these pellets and the results are discussed as follows. Pellets of hemin (nominally pure powders), $CoX_2(PPh_3)_2$ (nominally pure powders) and $NiCl_2(PPh_3)_2$ (nominally pure powders mixed with HDPE) were pressed to ~2 mm thickness in air with a manual hydraulic press (MTI) using a 13 mm stainless steel pellet die (Specac) with 3 tons of applied mass. Pellets of $Fe(H_2O)_6(BF_4)_2$ (nominally pure powders mixed with HDPE) were pressed to ~2 mm thickness in a $N_2$-filled glovebox (Innovative Technology) with a manual hydraulic press (Specac) using a 13 mm stainless steel pellet die (Specac) with 3 tons of applied mass.

Though the diameters of the pellets were 13 mm, the focused THz beam with a spot size of approximately ~5 mm diameter throughout the 2-mm samples defined the ~40 $mm^3$ sample volumes that were measured. The spin number density in the samples was on the order of $10^{23}$ $cm^{-3}$. The measured signals emerged from approximately $10^{20}$ spins within the spot of the focused THz beam.

**6. Powder X-ray diffraction**

Powder X-ray diffraction (PXRD) patterns were recorded with a Bruker D8 Advance diffractometer equipped with a Göbel mirror, rotating sample stage, LynxEye detector and Cu K$\alpha$ ($\lambda$ = 1.5405 Å) X-ray source in a $\theta/2\theta$ Bragg-Brentano geometry. An anti-scattering incident source slit (2 mm) and an exchangeable steckblende detector slit (8 mm) were used. The tube voltage and current were 40 kV and 40 mA, respectively. Knife-edge attachments were used to remove scattering at low angles. Samples for PXRD were prepared by placing a thin layer of the designated materials on a zero-background silicon (510) crystal plate.

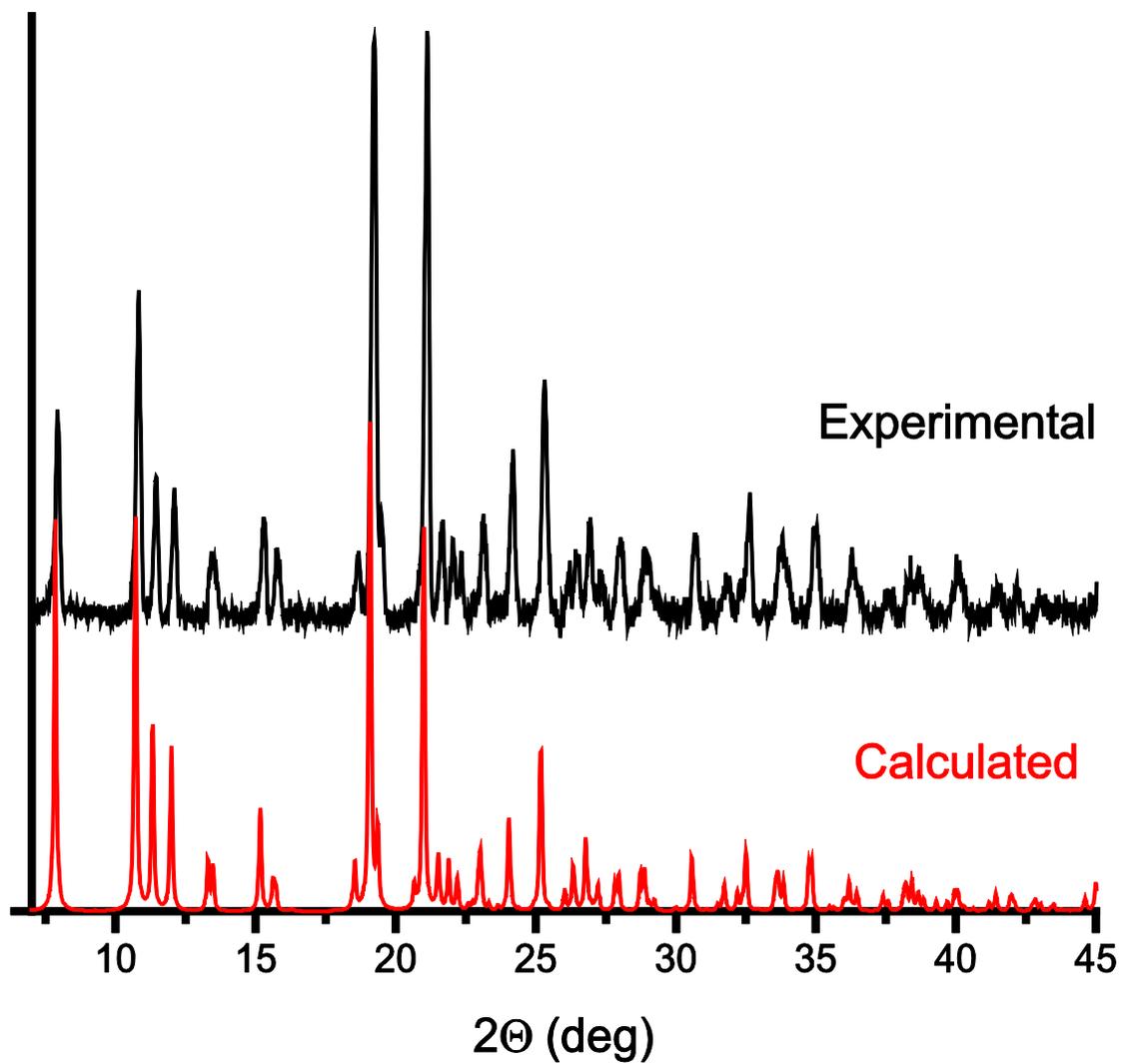

**Figure S13.** Background-corrected experimental (black) and calculated (red) PXRD patterns of $CoCl_2(PPh_3)_2$.

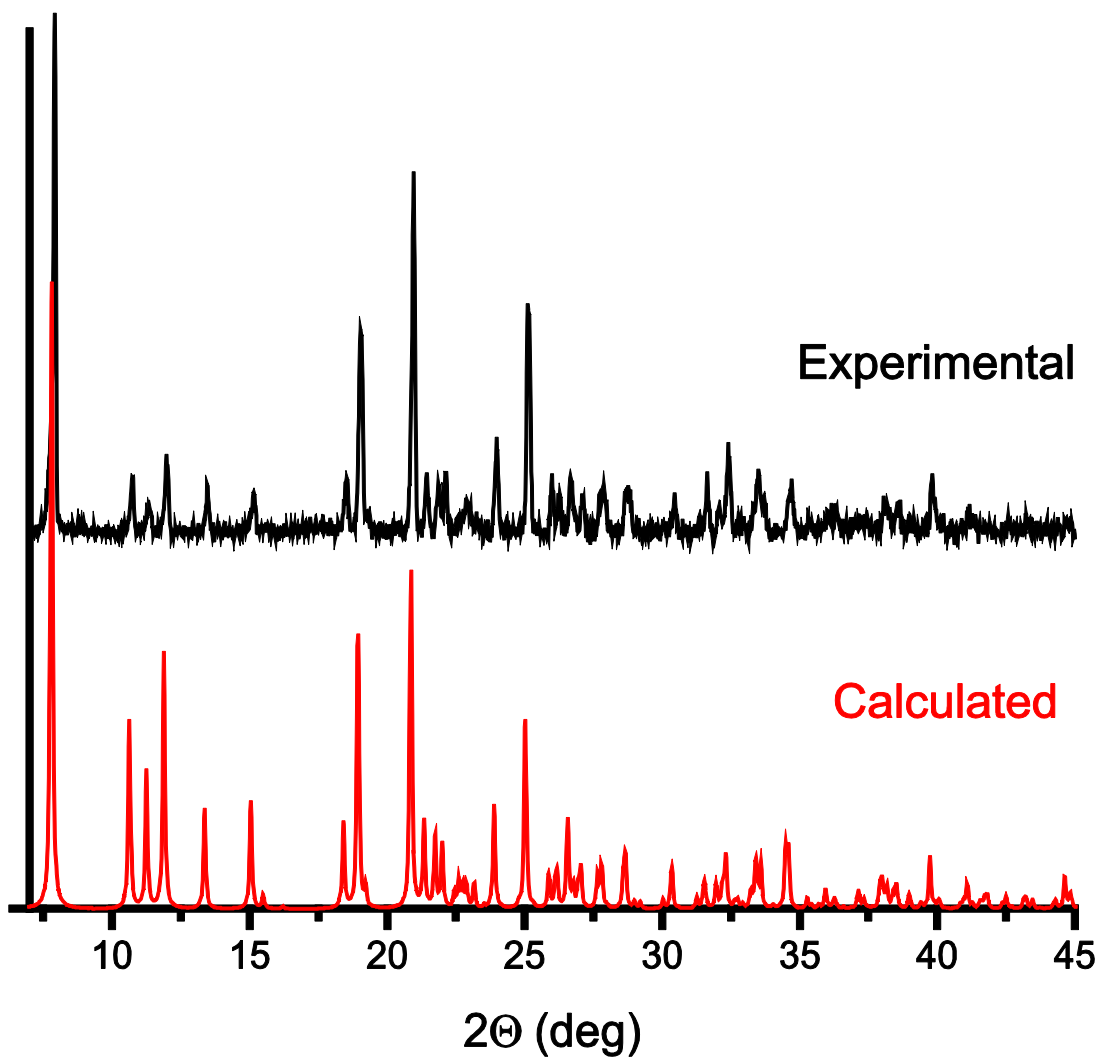

**Figure S14.** Background-corrected experimental (black) and calculated (red) PXRD patterns of CoBr$_2$(PPh$_3$)$_2$.